\pdfoutput=1 
\documentclass[11pt]{article}
\PassOptionsToPackage{table}{xcolor}

\usepackage{jheppub} 

\usepackage[T1]{fontenc} 
\usepackage{tangocolors}

\usepackage{tikz}
\usetikzlibrary{arrows,snakes,backgrounds}
\usetikzlibrary{decorations.pathmorphing,backgrounds,shapes,arrows,shadows}
\tikzset{zigzag/.style={decorate,decoration=zigzag}}
\tikzset{snake it/.style={decorate, decoration=snake}}
\makeatletter
\def\@hex@@Hex#1%
 {\if a#1A\else \if b#1B\else \if c#1C\else \if d#1D\else
  \if e#1E\else \if f#1F\else #1\fi\fi\fi\fi\fi\fi \@hex@Hex}
\makeatother

\usepackage[all]{xy}
\usepackage[percent]{overpic}
\usepackage{slashed}
\usepackage{wrapfig}
\usepackage{tabu}
\usepackage{diagbox}
\usepackage{mathrsfs,amsmath,amssymb,amsthm,amsfonts,tikz,graphicx,accents,hyperref, color}
\usepackage{dsfont,epiolmec, latexsym, stmaryrd, comment}
\usepackage{slashed,ccaption}
\usepackage{mathrsfs, calligra}
\usepackage{leftidx}
\usepackage{import}
\usepackage{multirow}
\usepackage{amsfonts}
\usepackage{pifont}
\usepackage{tabularx}
\usepackage[utf8]{inputenc}
\usetikzlibrary{intersections,calc}
\usepackage{ifthen}
\usepackage{amsmath}

\usepackage{caption} 
\DeclareMathOperator{\extdm}{d}
\newcommand{\extd}{\extdm \!}

\usepackage{array}

\usepackage{soul}

\hypersetup{ linktoc=all,
    colorlinks, linkcolor={palatinateblue},
    citecolor={brightpink}, urlcolor={amaranth}
}

\graphicspath{{Images/}}

\renewcommand{\d}[1]{\ensuremath{\operatorname{d}\!{#1}}}

\def\sideremark#1{\ifvmode\leavevmode\fi\vadjust{\vbox to0pt{\vss
 \hbox to 0pt{\hskip\hsize\hskip1em
 \vbox{\hsize2cm\tiny\raggedright\pretolerance10000
 \noindent #1\hfill}\hss}\vbox to8pt{\vfil}\vss}}}%
\DeclareSymbolFont{extraup}{U}{zavm}{m}{n}
\DeclareMathSymbol{\varheart}{\mathalpha}{extraup}{86}
\DeclareMathSymbol{\vardiamond}{\mathalpha}{extraup}{87}
\makeatletter
\renewcommand*{\@fnsymbol}[1]{\ensuremath{\ifcase#1\or \clubsuit \or \vardiamond \or \varheart\or
    \spadesuit\or \mathparagraph\or \|\or **\or \dagger\dagger
    \or \ddagger\ddagger \else\@ctrerr\fi}}
\makeatother

\definecolor{rosy}{RGB}{230,235,252}
\definecolor{myframetitle}{RGB}{90,89,170}
\definecolor{myblocktitle}{RGB}{140,185,249}
\definecolor{mytitle}{RGB}{10,80,26}

\definecolor{darkgreen}{RGB}{27,130,45}
\definecolor{darkblue}{rgb}{0,0,0.3}
\definecolor{darkred}{rgb}{0.7,0,0}

\definecolor{light gray}{RGB}{220,220,220}
\definecolor{dark purple}{RGB}{108,0,217}
\definecolor{pink}{RGB}{190,20,100}
\definecolor{orang}{RGB}{193,63,0}
\definecolor{green}{RGB}{11,98,17}
\definecolor{darkpink}{RGB}{153,0,76}
\definecolor{bluegreen}{RGB}{0,102,102}
\definecolor{greenlagan}{RGB}{0,102,0}
\definecolor{redgreen}{RGB}{102,102,0}
\definecolor{Redgreen}{RGB}{153,76,0}
\definecolor{vividviolet}{rgb}{0.62, 0.0, 1.0}
\definecolor{amaranth}{rgb}{0.9, 0.17, 0.31}
\definecolor{palatinateblue}{rgb}{0.15, 0.23, 0.89}
\definecolor{brightpink}{rgb}{1.0, 0.0, 0.5}
\definecolor{cornflowerblue}{rgb}{0.39, 0.58, 0.93}
\definecolor{deepcarminepink}{rgb}{0.94, 0.19, 0.22}
\definecolor{radicalred}{rgb}{1.0, 0.21, 0.37}

\newcommand\bc[1]{\boldsymbol{\mathcal{#1}}}

\newcommand{\Om}{{\Omega}} 

\DeclareFontFamily{OT1}{rsfs}{}

\DeclareFontShape{OT1}{rsfs}{m}{n}{ <-7> rsfs5 <7-10> rsfs7 <10->rsfs10}{} 

\DeclareMathAlphabet{\mycal}{OT1}{rsfs}{m}{n}

\newcommand{\be}{\begin{equation}}
\newcommand{\ee}{\end{equation}}
\makeatletter \@addtoreset{equation}{section}

\begin{document}


\preprint{TUW--21--03}

\newcommand{\mytitle}{
Null boundary phase space: slicings, news \& memory}

\title{\mytitle}

\author[a,b]{H.~Adami}
\author[c]{, D.~Grumiller}
\author[d]{, M.M.~Sheikh-Jabbari}
\author[d,e]{, V.~Taghiloo}
\author[b]{, H.~Yavartanoo}
\author[c,f]{and C.~Zwikel}
\affiliation{$^a$ Yau Mathematical Sciences Center, Tsinghua University, Beijing 100084, China}
\affiliation{$^b$ Beijing Institute of Mathematical Sciences and Applications (BIMSA), Huairou District, Beijing 101408, P. R. China}
\affiliation{$^c$ Institute for Theoretical Physics, TU Wien, Wiedner Hauptstrasse 8--10/136, A-1040 Vienna, Austria}
\affiliation{$^d$ School of Physics, Institute for Research in Fundamental
Sciences (IPM),\\ P.O.Box 19395-5531, Tehran, Iran}
\affiliation{$^e$ Department of Physics, Institute for Advanced Studies in Basic Sciences (IASBS),
P.O. Box 45137-66731, Zanjan, Iran}
\affiliation{$^f$ Perimeter Institute for Theoretical Physics, 31 Caroline Street North, Waterloo, Ontario, Canada N2L 2Y5}

\emailAdd{hamed.adami@bimsa.cn,  grumil@hep.itp.tuwien.ac.at, jabbari@theory.ipm.ac.ir,  v.taghiloo@iasbs.ac.ir, yavar@bimsa.cn, czwikel@perimeterinstitute.ca}

\abstract{
We construct the  boundary phase space in $D$-dimensional Einstein gravity with a generic given co-dimension one null surface ${\cal N}$ as the boundary. The associated boundary symmetry algebra is a semi-direct sum of diffeomorphisms of $\cal N$ and Weyl rescalings. It is generated by $D$ towers of surface charges that are generic functions over $\cal N$. These surface charges can be rendered integrable for appropriate slicings of the phase space, provided there is no graviton flux through $\cal N$. In one particular slicing of this type, the charge algebra is the direct sum of the Heisenberg algebra and diffeomorphisms of the transverse space, ${\cal N}_v$ for any fixed value of the advanced time $v$. Finally, we introduce null surface expansion- and spin-memories, and discuss associated memory effects that encode the passage of gravitational waves through $\cal N$, imprinted in a change of the surface charges.
}
\maketitle

\section{Introduction}

The study of field theories requires the specification of fall-off or boundary conditions, which can lead to physical degrees of freedom that reside at the boundary. In this work, we refer to them as boundary degrees of freedom (BDOF), to be distinguished from the usual bulk degrees of freedom, such as photon or graviton polarizations. In theories with local gauge invariance --- including gravitational theories --- BDOF are labeled and governed by specific gauge transformations that act non-trivially at the boundary, often called `non-proper gauge transformations'.

The number and type of BDOF depend on the precise boundary conditions. We are interested in maximizing the number of BDOF, in the sense that for a given bulk theory there exists no consistent set of boundary conditions that leads to more BDOF than this maximal number, for a given boundary. If such a set exists, then all other boundary conditions may be viewed as restrictions or deformations of such a maximal choice. Some of us argued in \cite{Grumiller:2020vvv} that such a set of maximal BDOF exists and made a specific proposal for it in $D$-dimensional Einstein gravity when the boundary is a given co-dimension one null surface ${\cal N}$: besides the $D(D-3)/2$ graviton  polarizations in the bulk, there are up to $D$ BDOF described by functions over ${\cal N}$. The quick counting works as follows: the metric has $D(D+1)/2$ independent components, of which $D(D-3)/2$ describe graviton polarizations. Of the difference, $2D$, half of the functions can be gauge fixed so that up to $D$ BDOF remain. Depending on the precise boundary conditions, some (or even all) of them can be pure gauge even at the boundary. 

A key question in this context that we address in the present work is how to conveniently construct, parametrize and label the maximal set of BDOF compatible with our assumptions about the boundary. We elaborate now a bit on what precisely we mean by `conveniently'. To do so, we recall a few basic technicalities.

A common method to label BDOF is to derive the surface charges associated with non-proper gauge transformations and diffeomorphisms, which may be computed, for instance, using the covariant phase space formalism \cite{Lee:1990nz, Iyer:1994ys, Compere:2018aar, Harlow:2019yfa}. For concreteness, we focus on the case of interest for the present work, $D$-dimensional Einstein gravity in presence of a boundary that is a co-dimension one null surface ${\cal N}$, though we expect many of our considerations generalize to gauge theories or Einstein gravity with matter and to timelike surfaces. There are several reasons why considering null surfaces as boundaries is of interest: they arise naturally in the asymptotic region of asymptotically flat spacetimes \cite{Bondi:1962,Sachs:1962,Barnich:2011ct}, in the near horizon region of black holes \cite{Donnay:2015abr,Afshar:2016wfy,Carlip:2017xne,Haco:2018ske}, and in the context of causal patch holography, see for instance \cite{Banks:2010tj,deBoer:2016pqk,Neiman:2017zdr} and refs.~therein. For $D<4$ such an analysis was carried through in \cite{Adami:2020ugu}, {see also \cite{Ruzziconi:2020wrb,Adami:2021sko,Geiller:2021vpg}}. For generic $D$ there are numerous earlier constructions, see e.g.~\cite{Donnay:2015abr,Hopfmuller:2016scf,Chandrasekaran:2018aop,Donnay:2019jiz,Grumiller:2019fmp,Jafari:2019bpw,Adami:2020amw,Fiorucci:2020xto,Chandrasekaran:2021hxc, Chen:2021kug,Freidel:2021fxf}, with a varying number of BDOF. 

In the present work, we construct and study the maximal set of BDOF for a given null hypersurface. This is achieved by solving the Einstein equations without imposing  boundary conditions, leading to a solution space involving $D(D-3)$ functions over $\cal N$ that correspond to the bulk gravitons and $D$ additional functions over $\cal N$ that specify the BDOF, in line with the analysis of \cite{Grumiller:2020vvv}. The covariant phase space formalism then establishes that this solution space is indeed a phase space with a well-defined symplectic structure. The solution space consists of the boundary phase space plus the bulk phase space.

The construction outlined above does not necessarily lead to a convenient organization of the BDOF. One key aspect is that the surface charges associated with non-proper diffeomorphisms can fail to be integrable on the field space or, equivalently, on the solution space. Physically, non-integrability of the surface charges is to be expected when bulk gravitons are allowed to have a non-vanishing flux through the boundary. The non-integrability is also closely related to non-conservation of these charges. The non-conservation is a consequence of having an open system, since the BDOF can interact non-trivially with themselves as well as with the bulk degrees of freedom. {This non-conservation} is captured by the null surface balance equation, which schematically is written as 
\begin{equation}
\frac{\extd}{\extd v}\,Q \sim - F
    \label{eq:angelinajolie}
\end{equation}
reviewed in more detail in the body of our paper. The left-hand side describes the change of the surface charge $Q$ as function of advanced time $v$ along the null surface $\cal N$. The right-hand side contains the flux through the null surface $\cal N$.

In practice, however, it can also happen that the surface charges are not integrable in the absence of any physical fluxes. As mentioned in \cite{Grumiller:2020vvv} and made explicit in \cite{Adami:2020ugu, Adami:2021sko, Ruzziconi:2020wrb, Geiller:2021vpg,Grumiller:2021cwg}, integrability of the surface charges depends on the slicing used to describe the boundary phase space. We are going to be more explicit about what we mean by `slicing' in the body of our paper. For now, the reader can think of a change of slicing as field dependent redefinition of the symmetry generators. 

In our work, we define the news to be the non-integrable part of the surface charges. It can be separated into `genuine news' and `fake news'. The former is news generated by a graviton flux in the bulk, while the latter is present even in the absence of such a flux. We call slicings without fake news `genuine slicings', meaning that the surface charges are integrable in the absence of genuine news.  So, when above we stated that we were interested in a `convenient' parametrization of the BDOF, technically we mean genuine slicings.

The conjecture put forward in \cite{Grumiller:2020vvv} and verified for some examples in \cite{Adami:2020ugu, Adami:2021sko,Ruzziconi:2020wrb, Geiller:2021vpg,Grumiller:2021cwg} states that there exist phase space slicings in which there are no fake news, and the non-integrable part of the surface charges is determined entirely by genuine news. In other words, the conjecture posits that there always exists at least one genuine slicing. In this work, we verify this conjecture for $D$-dimensional Einstein gravity (possibly with cosmological constant) with a null boundary $\cal N$. 

Having covered the existence of genuine slicings, it is natural to ponder about uniqueness. An important feature mentioned in \cite{Grumiller:2019fmp}, expanded more formally in \cite{Adami:2020ugu} and discussed for the example of topologically massive gravity in \cite{Adami:2021sko} is that genuine slicings are not unique and the surface charge algebra is slicing dependent. In particular, there exists a slicing, dubbed `Heisenberg slicing', in which the algebra associated with the boundary phase space takes the form of a direct sum of the Heisenberg algebra and diffeomorphisms on co-dimension two surfaces. In this work, we confirm that the same structure appears generically in $D$-dimensional Einstein gravity.

Besides confirming these expectations of earlier studies and generalizing them to arbitrary dimensions, we formulate null boundary memory effects. They arise when some bulk graviton flux passes through the null boundary $\cal N$. More specifically, we introduce two different kinds of memories {effects}, null surface expansion memory and null surface spin memory where the passage of a gravitational shockwave through the null boundary leaves an imprint on the surface charges. 

This paper is organized as follows. In section \ref{sec:metric-expansion-prelim} we set up the problem by choosing an adapted coordinate system around a generic null surface ${\cal N}$. In section \ref{sec:solution space} we impose the Einstein equations near the null boundary and construct the null boundary solution space. In section \ref{sec:NBS-generators} we explore null boundary symmetries, i.e., diffeomorphisms that keep intact the null boundary and move us within the associated solution space. In section \ref{sec:charges} we construct surface charge variations associated with the null boundary symmetries using the covariant phase space formalism and present the charge analysis in different slicings, the thermodynamic slicing, and a family of genuine slicings, in particular the Heisenberg slicing where the algebra of surface charges is a direct sum of Heisenberg algebra and  $D-2$ dimensional diffeomorphisms. In section \ref{sec:charge conservation balance equation} we study the (non-)conservation of our surface charges and the null surface balance equation \eqref{eq:angelinajolie} relating the time variation of the charges to the flux through the null boundary. In section \ref{sec:Integrable-cases} we discuss two physically relevant cases where the charges are integrable, namely when the null surface has vanishing expansion and when the graviton news through the null boundary vanishes. In section \ref{sec:GW-through-horizon} we introduce two types of null surface memory effects, expansion- and spin-memory. In particular, we study how our surface charges dynamically change when a gravitational wave passes through the horizon of a stationary black hole. Section \ref{sec:discussion} is devoted to concluding remarks. In appendix \ref{appen:null-boundary-EOM} we analyze the Einstein equations without expansion near the null boundary. In appendix \ref{appen:CPSF} we present a quick review of the covariant phase space formulation adapted for null boundaries and display the symplectic potential. In appendix \ref{appen:another-geuine-slicing} some additional genuine slicings of the  null boundary phase space are presented. In appendix \ref{appen:Kerr-Metric-GNC} we rewrite the Kerr solution in the coordinate system adopted here and discuss its conserved charges.

\section{General near null surface metric} \label{sec:metric-expansion-prelim}

Let $\mathcal{N}$ be a given smooth co-dimension one null hypersurface in a $D$ dimensional spacetime of signature $(-,+,\dots,+)$. In a neighborhood of any such hypersurface one can adopt Gaussian null-type coordinates that we set up as follows. Let $v$ be the advanced time coordinate along the null hypersurface such that the null surface is defined by
\begin{equation}
    g^{\mu \nu}\, \partial_\mu v\, \partial_\nu v =0\, .
\end{equation}    
A ray is defined as the vector tangent to this surface, $k^\mu = \eta\, g^{\mu \nu} \partial_\nu v$, where $\eta$ is an arbitrary non-zero function and $r$ the affine parameter of the generator $k^\mu$ such that $k^{\mu} = \d x^{\mu}/\d r=\delta^\mu_r$. The remaining $D-2$ coordinates $x^A$ are chosen as constants along each ray, $k^\mu \partial_\mu x^A =0$. These assumptions, while useful for numerous applications, come with some loss of generality and reduce the number of BDOF. We shall come back to generalizations and what they imply geometrically in the concluding section.

\begin{figure}[t]
\def \L {3.0}
    \centering
\begin{tikzpicture}
  \draw[thick,red] (-0.6*\L,-0.6*\L) coordinate (b) -- (0.9*\L,0.9*\L) coordinate (t);
  \draw[blue,->] (1.7*\L,-0.2*\L) -- (0.6*\L,0.9*\L);           
  \draw[blue,->] (1.6*\L,-0.3*\L) -- (0.4*\L,0.9*\L);           
  \draw[blue,->] (1.5*\L,-0.4*\L) -- (0.2*\L,0.9*\L);
  \draw[blue,->] (1.3*\L,-0.6*\L) -- (-0.2*\L,0.9*\L);
  \draw[blue,->] (1.2*\L,-0.7*\L) -- (-0.4*\L,0.9*\L);
  \draw[black,thick,->] (-0.1*\L,-0.1*\L)--(-0.09*\L,-0.09*\L); \draw[black,thick,->] (-0.08*\L,-0.08*\L)--(-0.07*\L,-0.07*\L);
  \draw[red] (-0.08*\L,0.04*\L) node[left, rotate=45] (scrip) {{$v$}};
  \draw[blue] (1.15*\L,-0.25*\L) node[left, rotate=-45] (scrip) {\small{ infalling null rays}};
  \draw[red] (-0.4*\L,-0.2*\L) node[left, rotate=45] (scrip) {\small{$r=0$}}; 
  \draw[red] (-0.3*\L,-0.45*\L) node[left, rotate=45] (scrip) {\small{${\cal N}$}};
  \draw[brown] (0.4*\L,-0.1*\L) node[left] (scrip) {\small{$r>0$}};
  \draw[brown] (0*\L,0.2*\L) node[left] (scrip) {\small{$r<0$}};
\end{tikzpicture}
\caption{Section of null hypersurface $\cal N$ at $r=0$ in $rv$-plane. Infalling null rays traverse $\cal N$ at different values of advanced time $v$. Each point on the red line corresponds to a transverse surface ${\cal N}_v$.
}\label{Fig:Null-surface-flux-Horizon}
\end{figure}
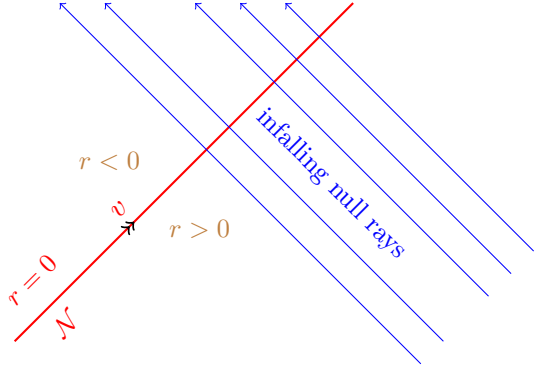

Without loss of generality, we take the null surface ${\cal N}$ to be localized at vanishing affine parameter, $r=0$, as depicted in Fig.~\ref{Fig:Null-surface-flux-Horizon}. The null surface ${\cal N}$ is assumed to have the topology  $\mathbb{R}_v\ltimes {\cal N}_v$, where ${\cal N}_v$ is the $D-2$ dimensional constant-$v$ subspace on ${\cal N}$ which is spanned by $x^A$.  We refer to ${\cal N}_v$ as  \textit{transverse surface}.\footnote{%
This transverse surface is sometimes called corner \cite{Donnelly:2016auv}. However, the latter terminology is used to develop a co-dimension two description of gravity while here we elaborate on a co-dimension one point of view. When describing future null infinity, the transverse surface is the celestial sphere \cite{Pasterski_2017,Donnay_2020}.}
In these adapted coordinates inverse metric and metric have the following vanishing components
\begin{equation}
    g^{vv}= 
    g^{vA} = 
    g_{rr}=
    g_{rA} =0 \,.
\end{equation}
The line-element 
\begin{equation}\label{G-F-M-01}
    \d s^2=  -V \d v^2 + 2 \eta \d v \d r + g_{{AB}} \left( \d x^A + U^A \d v\right) \left( \d x^B + U^B \d v\right)\, 
\end{equation}
depends on generic functions of all coordinates, $V, U^A, g_{{AB}}$, as well as on the function $\eta = \eta(v, x^A){>0}$. (Geodecity, $k\cdot \nabla k =0$, implies $\partial_r \eta =0$.\footnote{A null ray always satisfies the geodesic equation. Demanding that $r$ be an affine parameter along the ray implies that $\eta$ must be independent of $r$.})

We assume that the locus of the null surface, $r=0$, is not singular and that the metric coefficients admit a Taylor series expansion in powers of $r$ around $r=0$.
\begin{equation}\label{nearN-expansion}
            V=2\big(\eta\kappa-\mathcal{D}_v \eta \big)  r
           + \mathcal{O}(r^2)\,, \quad
        U^A= {\cal U}^A {- \frac{\eta}{\Omega}\Upsilon^A r} + \mathcal{O}(r^2)\,, \quad
        g_{{AB}}=\Om_{AB}- 2\eta\lambda_{AB}\, r
        + \mathcal{O}(r^2) 
\end{equation}
where all expansion coefficients are functions of $v,x^A$ and 
\be\label{OmegaAB-gammaAB}
\Omega:=\sqrt{\det \Omega_{AB}}\qquad\qquad \Omega_{AB}=\Omega^{2/(D-2)}\gamma_{{AB}} \qquad\qquad \det \gamma_{AB} =1\,,
\ee
where $\gamma_{AB}$ is an arbitrary unimodular matrix. To have a non-degenerate volume form, ${\sqrt{-\det{g_{\mu\nu}}}|_{r=0}=\eta \Omega}$, we assume {$\Omega,\eta>0$}. The function $\eta$ yields the volume of the $v,r$ part of the metric. We use the definition\footnote{
The transversal volume form $\Omega$ is a scalar density of weight $+1$, the quantity $\Upsilon^A$ is a vector density of weight $+1$, and the induced co-dimension two metric $\gamma_{{AB}}$ is a tensor-density of weight $-2/(D-2)$ in the $D-2$ dimensional sense.}
\be\label{Dv}
\mathcal{D}_v := \partial_v - \mathcal{L}_{\mathcal{U}}
\ee 
where $\mathcal{L}_{\mathcal{U}}$ is the Lie derivative along $\mathcal{U}^A$. As discussed in section \ref{sec:solution space}, the Einstein equations specify higher order coefficients in $r$ in terms of the leading order functions.

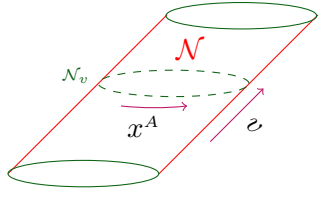
\begin{figure}
    \centering
\begin{tikzpicture}
\draw[red,rotate=-45] (0,0) -- (0,2.99);
\draw[red,rotate=-45] (1.42,1.414) -- (1.42,4.406);
\draw[green] (1,0) ellipse (28.5pt and 5pt);
\draw[green] (3.1,2.1) ellipse (28.5pt and 5pt);
\draw[green, style=dashed] (2.2,1.2) ellipse (28.5pt and 5pt);
\draw[pink,rotate=-45,->]  (1.6,2.2) -- (1.6,3.2);
\node[rotate=-45,right=] at (3.1,.8) {$\tiny{v}$};
\draw[pink,->] (1.5,.9) arc (240:300:25pt and 5pt);
\node[below=] at (1.8,.9) { $\tiny{x^{\text{\tiny $A$}}}$};
\node[right=] at (2.1,1.65) { {\textcolor{red}{$\mathcal{N}$}}};
\node[left=] at (1.22,1.3) { \tiny{\textcolor{green}{$\mathcal{N}_{v}$}}};
\end{tikzpicture}
  \caption{ Depiction of co-dimension one null boundary $\mathcal{N}$. ${\cal N}$ has the topology of $\mathbb{R}_v\ltimes {\cal N}_v$ where the transverse surface ${\cal N}_v$ is typically a $D-2$ dimensional spacelike compact surface. }
\label{fig:null-cylinder}
\end{figure}

To decompose the bulk metric adapted to  null hypersurfaces, it is standard to define two null vector fields $l^\mu, n^\mu$ ($l^2=n^2=0$) such that $l\cdot n=-1$,  $l^\mu$ is outward pointing and $n^\mu$ inward pointing. In adapted coordinates the associated 1-forms read
\begin{equation}\label{gennullbndryl}
    l :=  l_\mu \d x^\mu= -\frac{1}{2} V \d v  + \eta \d r \qquad \qquad
    n :=  n_\mu \d x^\mu= -\d v 
\end{equation}
{and the corresponding vector fields are given by}
\begin{equation}\label{l-n-vector}
    l^\mu \partial_\mu = \partial_v -U^A\partial_A + \frac{V}{2\eta} \partial_r \qquad\qquad n ^\mu \partial_\mu = - \frac{1}{\eta} \partial_r\,.
\end{equation}
From \eqref{l-n-vector} we see that ${\cal D}_v$ defined in \eqref{Dv} is the Lie derivative along the vector $l$ evaluated on ${\cal N}$. 
In terms of $l,n$, the induced co-dimension two metric
\begin{equation}\label{g-q-nl}
    q_{\mu \nu}= g_{\mu \nu}+ l_\mu n_\nu + l_\nu n_\mu \qquad\qquad q_{\mu \nu} l^\mu=q_{\mu \nu} n^\mu=0
\end{equation}
{yields the line-element on $\mathcal N$} 
\begin{equation}\label{metric-on-N}
   \d s^2_{\text{\tiny $\mathcal{N}$}}= \Omega_{AB} \left( \d x^A + \mathcal{U}^A \d v\right) \left( \d x^B + \mathcal{U}^B \d v\right)\, .
\end{equation}
As depicted in Fig.~\ref{fig:null-cylinder}, $\Omega_{AB}=\Omega_{AB}(v,x^A)$ {is the} metric over ${\cal N}_v$. The inverse of the $D-2$ dimensional metric $\Omega_{AB}$ is denoted by $\Omega^{AB}$, $\Omega^{AB}\Omega_{BC}=\delta^A_C$, and $A,B$ indices are raised or lowered by them.

The deviation tensors,
\begin{equation}\label{B-tensors-r=0}
        B^{\text{\tiny $l$}} _{\mu \nu} :=  \big(q^{\alpha}_{\mu} q^{\beta}_{\nu}\nabla_{\beta} l_{\alpha}\big)\big|_{r=0} \qquad\qquad
         B^{\text{\tiny $n$}} _{\mu \nu} := \big(q^{\alpha}_{\mu} q^{\beta}_{\nu}\nabla_{\beta} n_{\alpha}\big)\big|_{r=0} 
\end{equation}
provide a convenient parametrization. One can decompose {them} into trace (=expansion), symmetric trace-less (=shear) and anti-symmetric (=twist) parts,\footnote{%
For the shear of $l$ we have used the unusual notation $N_{\mu\nu}$, because as we shall show later this quantity is related to the flux (news) of gravitons through the null surface ${\cal N}$.} 
\begin{equation}\label{B-l-n-r=0}
        B^{^l}_{\mu \nu} =  \frac{1}{D-2}   \Theta_l\, q_{\mu \nu} + N_{\mu \nu} + \omega^{\text{\tiny $l$}} _{\mu \nu}\qquad\qquad
         B^{\text{\tiny $n$}} _{\mu \nu} = \frac{1}{D-2}   \Theta_{{n}}\, q_{\mu \nu} + 
         L_{\mu \nu} + \omega^{\text{\tiny $n$}} _{\mu \nu} \, .
\end{equation}
One can show that the twists $\omega^{\text{\tiny $l$}} _{\mu \nu}, \omega^{\text{\tiny $n$}} _{\mu \nu}$ are zero, the expansions
on $\mathcal{N}$ are
\begin{equation}\label{Theta-l-Theta-n-def}
\Theta_l = (q^{\mu \nu} \nabla _\mu l_\nu) \big|_{r=0} = \frac{{\cal D}_v \Omega}{\Omega}  =
\frac{\partial_v \Omega}{\Omega} - \bar{\nabla}_A \mathcal{U}^{A}
\qquad\quad\,
\Theta_n =(q^{\mu\nu}\nabla_\mu n_\nu)\big|_{r=0}  = \Omega^{AB}\lambda_{AB}
\end{equation}
and the shears are 
\begin{equation}\label{N-AB-L-AB-def}    
N_{AB}= \frac{1}{2}{\cal D}_v \Omega_{AB}- \frac{\Theta_l}{D-2}  \Omega_{AB} =\frac{1}{2}\Omega^{\frac2{D-2}} \mathcal{D}_v \gamma_{AB}\qquad\qquad
        L_{AB}   =\lambda_{AB}- \frac{\Theta_n}{D-2}  \Omega_{AB} 
\end{equation}
where $\bar{\nabla}_A$ is the $(D-2)$-dimensional covariant derivative {with respect to the} metric $\Omega_{AB}$ {and} $X_{(A}Y_{B)}:=(X_AY_B+X_B Y_A)/2$ denotes symmetrization of indices. (We note for later purposes that $N^{AB}=-\frac12 {\cal D}_v\Omega^{AB}-\frac1{D-2}\Theta_l \Omega^{AB}.$)

Regarding the expansions \eqref{Theta-l-Theta-n-def} and shears \eqref{N-AB-L-AB-def} two comments are in order. For stationary black holes with a bifurcate Killing horizon,  both expansions  vanish at the bifurcation surface. While it is immediate to see that $\Theta{_l}$ vanishes in this case,  in the coordinate system we have adopted $\Theta_n$ {is} non-zero. This is a well-known artifact of Eddington--Finkelstein type of coordinate systems, since the bifurcation surface lies at infinite advanced time in these coordinates. In all physically interesting situations, including black hole formation and evaporation, the bifurcation surface is absent anyhow and our coordinate system is adapted to describe such processes. Our second comment concerns the shear $N_{AB}$, which is proportional to the Lie derivative of the unimodular metric $\gamma_{AB}$ along $v$. We shall refer to this shear as `infalling graviton modes', but note that we are no expanding around any specific background $\Omega_{AB}$, so $N_{AB}$ need not be some small excitation. Indeed, in our charge analysis and discussion of memory effects we shall see that non-linear terms in $N_{AB}$ play an important role. By contrast, the shear $L_{AB}$ will not play a comparable role.

For later use we introduce the H$\grave{\text{a}}$ji$\check{\text{c}}$ek one-form 
\begin{equation}
{\cal H}_{A} := (q_A{}^\nu l_\lambda \nabla_\nu n^\lambda) \big|_{r=0}={\frac{\Upsilon_A}{2\Omega}} 
+\frac{\partial_{A}\eta}{2\eta}
\end{equation}
and the {scalar function} $\Gamma$, 
\begin{equation}\label{Gamma-def}
   \Gamma:=-2\kappa   +\frac{2}{D-2} \Theta{_l} + \frac{\mathcal{D}_v \eta}{\eta}
 \end{equation}
that appears in the expressions for the charges in later sections. Note that the scalar $\kappa$ appearing in the series expansion of $V$ in \eqref{nearN-expansion} is the non-affinity of the null hypersurface generator $l \cdot \nabla l^{\mu}:= \kappa\, l^{\mu}$ on $\mathcal{N}$.

\section{Null boundary solution space}\label{sec:solution space}

The near null surface metric to leading and next-to-leading order \eqref{nearN-expansion} is specified by $2+2(D-2)+(D-1)(D-2) = D(D-1)$ functions of $v,x^A$. The first counting refers to our original variables used in \eqref{nearN-expansion}, i.e., 2 scalars, $\kappa,\eta$, 2 co-dimension two vectors, ${\cal U}_A,\Upsilon_A$, and 2 co-dimension two symmetric 2-tensors, $\Omega_{AB},\lambda_{AB}$. In this section, we use these quantities as our building blocks, additionally splitting $\Omega_{AB}$ into conformal factor $\Omega$ and conformal class $\gamma_{AB}$, but use additionally the various composite quantities introduced in the previous section when convenient. The main goal of this section is to count the number of free functions available after imposing on-shell conditions, in order to get the number of bulk and boundary degrees of freedom. 

We analyze the Einstein equations (with arbitrary cosmological constant $\Lambda$)
\begin{equation}
    \mathcal{E}_{\mu\nu}:= R_{\mu \nu}- \frac{2 \Lambda}{D-2}\,  g_{\mu \nu}=0 
    \label{eq:lalapetz}
\end{equation}
in a Taylor-expansion around $r=0$. See appendix \ref{appen:null-boundary-EOM} for more details of the analysis and the construction of the phase space without invoking a perturbative expansion around $r=0$.

The Einstein equations \eqref{eq:lalapetz} may be decomposed in terms of the Raychaudhuri equation $\mathcal{E}_{ll}=l^{\mu}l^{\nu}\mathcal{E}_{\mu\nu}=0$,  the Damour equation $\mathcal{E}_{l A}=l^{\mu}q_{A}{}^{\nu}\mathcal{E}_{\mu \nu}=0$ and the trace and trace-less parts of $\mathcal{E}_{AB}=0$. At zeroth order in $r$, they respectively lead to
\begin{subequations}\label{EoM-r0}
\begin{align}
        &{\cal D}_v  \Theta_l-\kappa  \Theta_l+\frac{1}{D-2} \Theta_l^2+N_{AB}N^{AB}=0\label{EoM-Raychaudhuri}\\
        &{\cal D}_v \Big(\Upsilon_{A}+\Omega \frac{\partial_A\eta}{\eta}\Big)-2\Omega\partial_{A}\Big( {\kappa +\frac{D-3}{D-2}\Theta_l}  \Big)+2 \Omega\bar{\nabla}^{B}N_{A B}=0
        \label{EoM-Damour}\\
        &{\cal D}_v  \Theta_{n}+\kappa  \Theta_{n}+\Theta_l\Theta_{n}-\big(\bar{\nabla}_{C}\mathcal{H}^{C}+\mathcal{H}^{C}\mathcal{H}_{C}\big)
+\frac12 \bar{R}-\Lambda=0\label{Thetan-EoM}\\
        &2\mathcal{D}_{v}L_{AB}-4{L_{(A}}^{C}{N_{B)C}}+\Theta_{n}N_{AB}+\Big(2\kappa+\frac{D-6}{D-2}\Theta_l\Big)L_{AB}+\bar{R}_{AB}\nonumber\\
    &-2\mathcal{H}_{A}\mathcal{H}_{B}-2\bar{\nabla}_{(A}\mathcal{H}_{B)}+\Big(2\bar{\nabla}_{C}\mathcal{H}^{C}+2\mathcal{H}^{C}\mathcal{H}_{C}-\bar{R}\Big)\frac{\Omega_{AB}}{D-2}=0 \label{trace-less-AB}
\end{align}
\end{subequations}
where ${\cal D}_v$ defined in \eqref{Dv} implicitly contains the vector ${\cal U}_A$, and $\bar{R}_{AB}$ is the intrinsic Ricci tensor of the co-dimension two metric $\Omega_{AB}$.  

The $D(D-1)/2$ equations above are dynamical as they involve $v$-derivatives. Alternatively, one may view \eqref{EoM-Raychaudhuri} and \eqref{EoM-Damour} as $D-1$ non-differential (in $v$) equations for $\kappa$ and ${\cal U}^A$ in terms of the other functions (and their $v$-derivatives). The last two equations, \eqref{Thetan-EoM} and \eqref{trace-less-AB} are first order $v$-derivative equations for $\lambda_{AB}$ and specify it up to $(D-1)(D-2)/2$ functions over ${\cal N}_v$. We denote these functions by $\hat \lambda_{AB}(x^A)$.

The remaining Einstein equations, $\mathcal{E}_{n n}=n^{\mu}n^{\nu}\mathcal{E}_{\mu\nu}, \mathcal{E}_{l n}=l^{\mu}n^{\nu}\mathcal{E}_{\mu\nu}, \mathcal{E}_{n A}=n^{\mu}\mathcal{E}_{\mu A}$ are, respectively, algebraic equations for the order $r^2$ terms in the expansion of the trace of $g_{AB}$, $\Omega^{AB}g_{AB}$, $V$, and $U^A$, and specify these higher order terms through lower order ones. Since the higher order terms do not appear in the analysis of symmetries and charges we do not display them. 

Even though it is not required for the charges, it is instructive to explore the Einstein equations to higher order in $r$. Again, $\mathcal{E}_{n n}, \mathcal{E}_{n l}, \mathcal{E}_{l l}, \mathcal{E}_{l A}, \mathcal{E}_{n A}$ determine higher order terms in the expansion of $\Omega^{AB}g_{AB}, V, U^A$, whereas $\mathcal{E}_{AB}$ yield equations for higher order terms in the traceless parts of $g_{AB}$ and $\hat{\lambda}^{(n)}_{AB}$. These are first order differential equations in $v$ and hence determine $\hat{\lambda}^{(n)}_{AB}$ up to functions over ${\cal N}_v$, $\hat{\lambda}^{(n)}_{AB}(x^A)$. One may resum them into a single function at a constant $v$ surface as $\hat{g}^{(v)}_{{AB}}(r, x^A):=\sum_{n=1}^\infty \hat{\lambda}^{(n)}_{AB}(x^A) r^n$, where $\hat{\lambda}^{(1)}_{AB}=\hat \lambda_{AB}(x^A)$. 

To specify a solution in our null boundary solution space one should give $D+D(D-3)+1$ functions over $\cal N$. This number is just the difference between the original number of free functions, $D(D-1)$, and the number of non-differential (in $v$) equations that determine $\kappa$ and ${\cal U}_A$. The first $D$ of these functions are $\eta, \Omega, {\Upsilon}^A$. As we shall demonstrate in the next sections, these functions feature in the boundary charges and thus can be associated with BDOF. The $D(D-3)$ functions correspond to $\gamma_{AB}$ and the traceless part of $\hat{g}^{(v)}_{{AB}}$, and constitute the bulk degrees of freedom --- from a Lagrangian perspective this number corresponds to the usual $D(D-3)/2$ gravitational wave helicities. Finally, the remaining $1$ function is $\Theta_n$, which in our construction does not constitute a degree of freedom. We shall come back to it in the concluding section. 

In summary, our analysis of this section shows that we have $D$ BDOF in addition to the usual bulk degrees of freedom.

\section{Null boundary symmetries}\label{sec:NBS-generators}

We analyze the diffeomorphisms that preserve our null boundary structure in section \ref{sec:4.1} and then determine their algebra in section \ref{sec:4.2}.

\subsection{Null boundary preserving diffeomorphisms}\label{sec:4.1}

Diffeomorphisms generated by the vector field
\begin{multline}\label{NBS-vector}
\xi = T\,\partial_v + \Big( r(\mathcal{D}_{v}T-W)
      -r^2 \frac\eta2 \Big(\frac{\Upsilon_A}\Omega - \frac{\partial_A \eta}{\eta}\Big) \partial^{A}T
        +\mathcal{O}(r^3)\Big)\,\partial_r \\
        + \Big(Y^A -r \eta  \partial^A T {-}r^2{\eta^2} \lambda^{AB}\partial_{B}T + \mathcal{O}(r^3)\Big)\,\partial_A
\end{multline}
keep $r=0$ as a null surface, where $T=T(v,x^A)$, $W=W(v,x^A)$ and $Y^A=Y^A(v,x^A)$ are the symmetry generators. Since the Einstein equations are covariant, these diffeomorphisms move us in the solution space constructed in the previous section, namely 
\begin{subequations}
    \begin{align}
       &{\delta_\xi \eta = 2\eta {\cal D}_v T + T  \partial_{v}\eta -W\eta  +Y^A \partial_A \eta}\\
        &\delta_\xi\Omega=T \Omega \Theta_l + \Omega \bar{\nabla}_{A}(Y^{A}+\mathcal{U}^A T)\\ 
       &\delta_\xi \mathcal{U}^{A} ={\cal D}_v (Y^A + T \mathcal{U}^{A})\\
       & \delta_\xi \Omega_{AB} = \mathcal{L}_{(Y+T\mathcal{U})} \Omega_{AB} +\frac{2}{D-2}\, T \Theta_l \Omega_{AB}+ 2 T N_{AB} \\
       &{\delta_{\xi}\kappa= {\cal D}_v({\cal D}_vT+ T\kappa )+(Y^{A}+\mathcal{U}^A T)\partial_{A}\kappa} \\
      & {\delta_{\xi}\Theta_l = \mathcal{D}_v (T \Theta_l)+(Y^A + T \mathcal{U}^{A})\partial_A \Theta_l }\\
     &{\delta_{\xi}\Upsilon_{A}= T {\cal D}_v \Upsilon_A+\mathcal{L}_{(Y+T \mathcal{U})}\Upsilon_{A}+\Omega (\partial_{A}W- {\Gamma}\partial_{A}T-{2}{N_{A B}}\partial^{B}T)}\\
            &{\delta_{\xi}\Gamma=-{\cal D}_{v}(W-\Gamma T)+(Y^{A}+\mathcal{U}^A T)\partial_A\Gamma} \\
    &\delta_{\xi}N_{AB}=\mathcal{D}_{v}(TN_{AB})+\mathcal{L}_{(Y+T\mathcal{U})}N_{AB}\label{N-AB-transf}\\
          & {\delta_\xi\lambda_{AB}= \mathcal{D}_{v} (T\lambda_{AB})+\mathcal{L}_{(Y+T{\cal U})}\lambda_{AB}-2\lambda_{AB}\mathcal{D}_{v}T+2\left(\mathcal{H}_{(A}+\bar{\nabla}_{(A}\right)\bar{\nabla}_{B)}T}
    \end{align}
\end{subequations}
were ${\cal L}_Y$ denotes the Lie derivative along $Y^A$. 

The above transformations, when acting on different functions, can be homogeneous or inhomogeneous. The homogeneous ones are those that remain zero under transformations if they are zero at some point in the solution space. For example, $\Theta_l$ and $N_{AB}$ transform homogeneously. On the other hand, functions such as $\kappa$, $\mathcal{U}^A$ and $\Upsilon_A$ transform inhomogeneously under the diffeomorphisms \eqref{NBS-vector}.


\subsection{Algebra of null boundary symmetries}\label{sec:4.2}

Using the adjusted Lie bracket\footnote{%
In computing the Lie bracket of symmetry generators associated with diffeomorphisms that depend on functions in the solution space, one should adjust for the field dependence and subtract the changes in the diffeomorphisms due to the change in the fields, viz., $ [\xi_1, \xi_2]_{{\textrm{\tiny adj. bracket}}}
=[\xi_1, \xi_2]-\delta_{\xi_1}\xi_2+\delta_{\xi_2}\xi_1$. This bracket was originally called ``modified Lie bracket'' in \cite{Barnich:2011mi}. However, as discussed in \cite{Compere:2015knw} the name adjusted bracket seems more appropriate. }
we have
\begin{equation}\label{3d-NBS-KV-algebra}
    [\xi(  T_1, W_1, Y_1^{A}), \xi( T_2,  W_2, Y_2^{A})]_{{\text{adj. bracket}}}=\xi(  T_{12}, W_{12}, Y_{12}^{A})
\end{equation}
where 
\begin{subequations}\label{W12-T12-Y12}
\begin{align}
    &T_{12}=\left(T_{1}\partial_{v}+Y_{1}^{A}\partial_{A}\right)T_{2}-(1\leftrightarrow 2),\\
    &W_{12}=\left(T_{1}\partial_{v}+Y_{1}^{A}\partial_{A}\right)W_{2}-(1\leftrightarrow 2),\\
    &Y_{12}^{B}=\left(T_{1}\partial_{v}+Y_{1}^{A}\partial_{A}\right)Y_{2}^{B}-(1\leftrightarrow 2).
\end{align}
\end{subequations}
The above algebra is Diff(${\cal N}) \inplus$ Weyl$({\cal N})$, where  Diff(${\cal N})$ is generated by $T,Y^A$ and Weyl$({\cal N})$ which denotes the Weyl scaling on ${\cal N}$, is generated by $W$. We refer to it as null boundary symmetry algebra.

The null boundary symmetry algebra Diff(${\cal N}) \inplus$ Weyl$({\cal N})$ has several interesting subalgebras. If we turn off $Y^A$ and $W$ sectors, the generator $T$ forms a Witt algebra (diffeomorphisms along $v$ direction) but with an arbitrary dependence in $x^A$. These generators were called ``T-Witt'' \cite{Adami:2020amw}. Turning off $T, W$ sectors, $Y^A$ generate diffeomorphisms of the transverse surface ${\cal N}_v$. Nonetheless, one should note that these diffeomorphisms have arbitrary $v$ dependence. A class of subalgebras arise from the fact that our generators are generic functions of $v$. If the $v$ direction has no special points, one may Taylor-expand the generators around any given point $v_0$ and keep terms up to the order that still close the algebra. As an example, consider the subalgebra  obtained through the following truncation
\begin{equation}\label{Freidel-algebra}
     T =t_0  + t_1 v + t_2 v^2\qquad\qquad     W =  w_0\qquad \qquad Y^A = y_0^A
\end{equation}
where $t_0, t_1, t_2; w_0, y^A_0$ are only function of $x^A$. The $t_i$ form an $sl(2,\mathbb{R})$ algebra and $w_0$ an abelian $u(1)$ algebra, Weyl$({\cal N}_v)$. This subalgebra is hence (Diff(${\cal N}_v) \inplus sl(2,\mathbb{R})_{{\cal N}_v}$)$\inplus$ Weyl$({\cal N}_v)$, which is closely related to the corner algebra discussed in \cite{Ciambelli:2021vnn, Freidel:2021cbc}. To be more precise, the algebra without the Weyl$({\cal N}_v)$ part was called corner symmetry algebra and the one which also includes the translations in $r$, $r\to r+R(x^A)$, was called extended corner algebra. In our case we do not have the latter, as we keep $r=0$ a null surface throughout.

\section{Surface charge analysis}\label{sec:charges}

The surface charge variation\footnote{%
See appendix \ref{appen:CPSF} for a short review of the covariant phase space method used to derive this result.}  associated with a symmetry generator $\xi$
\begin{equation}\label{surface-charge}
        \slashed{\delta} Q_\xi := \oint_{\partial \Sigma} \mathcal{Q}_\xi^{\mu \nu} \d x_{\mu \nu}
\end{equation}
expands in Einstein gravity as
\begin{equation}
    \mathcal{Q}_\xi^{\mu \nu} =\frac{\sqrt{-g}}{8 \pi G}\, \Big( h^{\lambda [ \mu} \nabla _{\lambda} \xi^{\nu]} - \xi^{\lambda} \nabla^{[\mu} h^{\nu]}_{\lambda} - \frac{1}{2} h \nabla ^{[\mu} \xi^{\nu]} + \xi^{[\mu} \nabla _{\lambda} h^{\nu] \lambda} - \xi^{[\mu} \nabla^{\nu]}h \Big)
\end{equation}
where $h_{\mu \nu}= \delta g_{\mu \nu}$, $h= g^{\mu \nu}\delta g_{\mu \nu}$, and $\partial \Sigma $ corresponds to the transverse surface ${\cal N}_v$. See appendix \ref{appen:CPSF} for more details.

Plugging \eqref{G-F-M-01} and \eqref{NBS-vector} into \eqref{surface-charge}, yields the surface charge variation
\begin{equation}\label{surface-charge-01}
        \slashed{\delta} Q_{\xi}= \frac{1}{16\pi G} \int_{{\cal N}_v} \d{}^{D-2} x \left(  W\delta\Omega+Y^{A}\delta\Upsilon_{A} + T \slashed{\delta} \mathcal{A}\right) 
\end{equation}
with
\begin{equation}
    \slashed{\delta} \mathcal{A}=-2 \Omega \delta \Theta_l +\Omega\Theta_l\frac{\delta\eta}{\eta}-  \Gamma \delta\Omega
+ \mathcal{U}^{A}\delta \Upsilon_{A}-\Omega N^{AB} \delta\Omega_{AB}\, .
\end{equation}
The notation $\slashed{\delta}$ is used to stress that the charge variation is not necessarily integrable in field space. Tackling the question of whether or not the charges are integrable requires specifying which combinations of the symmetry generators are taken to be field independent, which amounts to a choice of slicing of the phase space. 

By ``slicing'' we mean 
a specific choice of the field dependence of the symmetry generators (including, possibly, the choice that there is none). Changing the slicing means that one takes symmetry generators to a linear combination thereof while allowing for these coefficients to have general dependence on the fields in the solution space. 
Thus, there is no reason to consider no field dependence of the symmetry generators as more natural than some other choice, since this notion is not even well-defined. 

In such change of slicing one keeps the same bulk theory with the same fall-off conditions, but relabels the state-dependence through redefinitions of the symmetry generators. Thus, one still describes the same phase space but it is reorganized/sliced differently. Inequivalent slicings in general will lead to inequivalent symmetry algebras, see section 4 of \cite{Adami:2020ugu} for more concise formulation of generic change of slicing (which was called change of basis in that work). The differences can be substantial, in the sense that central extensions, non-linearities and/or non-integrability may appear in one set of slicings but not in other sets of slicings, or even a Lie algebra of surface charges may be mapped onto an algebra which is not of the form of a Lie algebra, e.g.~see the example of Heisenberg-type algebra in \cite{Grumiller:2019fmp}. It is thus relevant to find the most suitable (classes of) slicings for a given physical setup. We shall present pertinent examples below, when discussing differences between thermodynamical and Heisenberg slicings.

Only after a slicing is specified, one can state whether or not the charges are integrable for this particular slicing. This implies that integrability of the charges is not solely a property of the bulk theory or the boundary conditions, but additionally may depend on the choice how to slice the phase space.

Physically, non-integrable charges are typically related to a non-vanishing flux through the boundary \cite{Wald:1999wa,Barnich:2011mi}, see more details on this in section \ref{sec:charge conservation balance equation}. Generally, $\slashed{\delta} Q$ is non-integrable over our null boundary solution space since we allow for fluxes through the boundary ${\cal N}$. This feature prevents us from working with the Poisson bracket of the charges. We use instead the modified bracket (MB) proposed by Barnich and Troessaert \cite{Barnich:2011mi},
\begin{subequations}\label{BT-Bracket}
    \begin{align}
   &\delta_{\xi_{2}}Q^{\text{I}}_{\xi_{1}} := \left\{Q^{\text{I}}_{{\xi_{1}}},Q^{\text{I}}_{{\xi_{2}}}\right\}_{{\tiny{\text{MB}}}} {-} F_{\xi_{2}}(\delta_{\xi_{1}}g) \label{BT-Bracket-01}\\
     &\left\{Q^{\text{I}}_{\xi_{1}},Q^{\text{I}}_{\xi_{2}}\right\}_{{\tiny{\text{MB}}}} =\, Q^{\text{I}}_{[\xi_{1},\xi_{2}]_{{\text{adj. bracket}}}}+K_{\xi_1,\xi_2}
     \label{BT-Bracket-02}
\end{align}
\end{subequations}
where $K_{\xi_1,\xi_2}$ is the central term, $Q^{\text{I}}_{\xi}$ the integrable part of the charges  and $F_{\xi}(\delta g)$ the non-integrable part, $ \slashed{\delta} Q_{\xi}=\delta Q^{\text{I}}_{\xi}+F_{\xi}(\delta g)$. The flux term is not necessarily antisymmetric $F_{\xi_{2}}(\delta_{\xi_{1}}g)\neq -F_{\xi_{1}}(\delta_{\xi_{2}}g)$, which we shall see more explicitly in examples below. 

The split into integrable and non-integrable parts is ambiguous and leads to a shift-ambiguity in the central term $K_{\xi_1,\xi_2}$ \cite{Barnich:2011mi}. To partially fix this ambiguity, we require the central term $K_{\xi_1,\xi_2}$ to be state independent, by which we mean that is does not vary over the solution space, see e.g.~section 5.1 of \cite{Adami:2020amw} for a more detailed discussion. 

An important aspect discussed, e.g., in \cite{Adami:2020ugu, Adami:2021sko, Ruzziconi:2020wrb, Geiller:2021vpg} is that the integrability of the charges and the presence or absence of fluxes do depend on the slicing. In the following, to shed new light on this issue, we discuss two classes of slicings. 

The first one, studied in section \ref{sec:thermoslicing}, is dubbed ``thermodynamic slicing''. In this slicing, $W,T,Y^A$ are state independent ($\delta W = \delta T=\delta Y^A=0$). This name will be justified in section \ref{sec:balance-eq-thermodynamic}, see also \cite{Adami:2021kvx}. The second one is a specific ``genuine slicing''. By this we mean any slicing in which the charges are integrable in the absence of bulk fluxes through the boundary, i.e., when there is no physical radiation through the boundary \cite{Grumiller:2020vvv, Adami:2020ugu, Adami:2021sko}.

\subsection{Thermodynamical slicing}\label{sec:thermoslicing}

The thermodynamic slicing is defined by state-independence of $W,T,Y^A$ in the vector field \eqref{NBS-vector}, $\delta W = \delta T=\delta Y^A=0$.

Applying the MB method discussed above and separating the integrable and flux parts, $\slashed{\delta}{Q}_\xi = \delta Q^{\text{I}}_\xi +F_\xi (\delta g)$, yields the integrable part
\begin{equation}\label{Integrable-part-Charge-Geometric}
    Q^{\text{I}}_\xi = \frac{1}{16\pi G} \int_{{\cal N}_v} \d{}^{D-2} x \big(  W\ \Omega+Y^{A}\ \Upsilon_{A}+ T\ ( -\Gamma \Omega + \mathcal{U}^{A}\Upsilon_{A}) \big)
\end{equation}
and the flux
\begin{equation}
\label{Non-Integrable-part-Charge-Geometric}
       F_{\xi}(\delta g ; g)= \frac{1}{16\pi G} \int_{{\cal N}_v} \d{}^{D-2} x \, T \Big(  -2\Omega \delta \Theta_l + \Omega \Theta_l \frac{\delta \eta}{\eta}+\Omega \delta \Gamma -\Upsilon_{A}\delta \mathcal{U}^{A} -\Omega N^{AB} \delta\Omega_{AB} \Big) \,.
\end{equation}
Straightforward but long computations show that the integrable part of the charges \eqref{Integrable-part-Charge-Geometric} satisfy the same algebra as the symmetry generators \eqref{3d-NBS-KV-algebra}, \eqref{W12-T12-Y12}, {i.e. Diff(${\cal N}) \inplus$ Weyl$({\cal N})$}. In particular, there is no central extension.  Explicitly, if we denote the charges associated with the symmetry generators $\xi(T,0,0)$, $\xi(0,W,0)$ and $\xi(0,0,Y^A)$ by  $\bc{T}(T), \bc{W}(W)$ and $\bc{J}(Y^A)$, respectively, then the MB bracket algebra reads
\begin{subequations}\label{Thermodynamic-slicing-algebra}
    \begin{align}
   \{\bc{T}(T_1), \bc{T}(T_2)\}_{_{\text{MB}}} &= \bc{T}(T_1\partial_v T_2-T_2\partial_v T_1),  \\
      \{\bc{J}(Y_1^A), \bc{J}(Y_2^B)\}_{_{\text{MB}}}&= \bc{J}(Y_1^A\partial_A Y_2^B-Y_2^A\partial_A Y_1^B),  \\
         {\{\bc{T}(T), \bc{J}(Y^A)\}_{_{\text{MB}}}}&= -\bc{T}(Y^A\partial_A T)+ \bc{J}(T\partial_v Y^A), \\
             \{\bc{W}(W_1), \bc{W}(W_2)\}_{_{\text{MB}}} &= 0, \\ 
      \{ \bc{T}(T), \bc{W}(W)\}_{_{\text{MB}}} &= \bc{W}(T\partial_v W), \\ 
        \{\bc{W}(W), \bc{J}(Y^A)\}_{_{\text{MB}}} &=  -\bc{W}(Y^A\partial_A W)\, .
     \end{align}
\end{subequations}

Consistently, in the absence of flux of bulk gravitons, $N_{AB}=0$, and in co-rotating frame, $\mathcal{U}^A =0$, we recover the results of \cite{Adami:2020amw}. For $D=3$, where the news tensor identically vanishes, one recovers the results obtained in Appendix C of \cite{Adami:2020ugu}. Moreover, as seen explicitly above, the MB procedure yields a vanishing central charge.

We close this section by justifying the name thermodynamic slicing. The zero mode charges associated with symmetry generators $\partial_v, -r\partial_r, \partial_A$, respectively, $\bc{T}(1), \bc{W}(1), \bc{J}(1)$, recover the usual thermodynamic charges if ${\cal N}$ is Killing of horizon of a black hole. Explicitly, $\bc{T}(1)$ corresponds to energy, $\bc{W}(1)$ to entropy and $\bc{J}(1)$ to angular momentum. These charges commute with each other; moreover, entropy commutes with all other charges. These points will be discussed in more detail in section  \ref{sec:balance-eq-thermodynamic}; see also \cite{Adami:2021kvx} for more elaborations.

\subsection{Genuine and Heisenberg  slicing}\label{sec:genuine-slicing}

The expression of the flux in the thermodynamic slicing \eqref{Non-Integrable-part-Charge-Geometric} is non-zero even in the absence of a graviton flux encoded in the tensor $N_{AB}$. As discussed in \cite{Adami:2020ugu, Adami:2021sko}, this flux depends on the slicing and one would expect that there should exist genuine slicings such that the flux is manifestly zero for vanishing genuine flux, by which we mean $N_{AB}=0$.

In this section, we present a one-parameter family of genuine slicings with the following property: its symmetry algebra at each $v$ has the structure of a direct sum of the symmetries of the transverse surface ${\cal N}_v$ and the symmetries normal to ${\cal N}_v$. This slicing is hence a direct-sum genuine slicing.  In particular, there is one member in this family such that the algebra is the direct sum of Diff(${\cal N}_v$) and Heisenberg algebra. This is referred to as the Heisenberg slicing.  Reaching such a slicing can be tedious {and one may first construct  an intermediate slicing in which the algebra has the form of semi-direct sum of Heisenberg and Diff(${\cal N}_v$) algebra.} This intermediate  genuine slicing as well as another example is presented in appendix \ref{appen:another-geuine-slicing}.

\paragraph{Direct-sum genuine slicings.} Starting from the thermodynamic slicing, consider {a one-parameter} family change of slicings
\begin{subequations}\label{tilde-slicing-generic-s}
    \begin{align}
    \tilde{W}&=W-\Gamma T-\left({Y}^A +T \mathcal{U}^A\right)\bar{\nabla}_A\mathcal{P}, \\ \tilde{T}^{(s)}&=e^{-s \mathcal{P}}\Omega\Theta_l  T+e^{-s \mathcal{P}}\bar{\nabla}_A(\Omega ({Y}^A +T \mathcal{U}^A))\\  \tilde{Y}^{A}&={Y}^A +T \mathcal{U}^A
\end{align}    
\end{subequations}
where $s$ is a real number and 
\begin{equation}\label{cal-P--def}
    \mathcal{P}:=\ln{\frac{\eta}{\Theta_l^2}}\,.
\end{equation}
As we see the change of slicing \eqref{tilde-slicing-generic-s} takes the original symmetry generators to a linear combination thereof with coefficients which depend on the fields  on the solution space and their derivatives. The change of slicing then amounts to taking $\delta\tilde W=\delta\tilde{T}^{(s)}=0=\delta\tilde{Y}^A$. Therefore,  the original symmetry generators, $W, T, Y^A$ have  non-zero variations in the new slicing, which is dictated by the requirement of new tilde-generators to have vanishing variations over the solution space. As a result the charges transform to a certain (in general non-linear) combination of the original charges \cite{Adami:2020ugu}.

The charge variation can be written as $\slashed{\delta}{Q}_{\xi}= \delta \tilde{Q}^{\text{I}}_{\xi} + \tilde{F}_\xi (\delta g)$, with the integrable part
\begin{equation}
     \tilde{Q}^{\text{I}}_{\xi}= \frac{1}{16\pi G} \int_{{\cal N}_v} \d{}^{D-2} x \left( \tilde{W} \Omega+\tilde{Y}^{A}  \mathcal{J}_{A}+\tilde{T}^{(s)} \mathcal{P}_{(s)}\right)
\end{equation}
and the flux
\begin{equation}
    \tilde{F}_\xi (\delta g)= -\frac{1}{16\pi G} \int_{{\cal N}_v} \d{}^{D-2} x\  \left[ e^{s\mathcal{P}}\tilde{T}^{(s)}-\bar{\nabla}_{C}(\Omega \tilde{Y} ^{C})\right]\Theta_l^{-1} N^{AB}\delta\Omega_{AB}
\end{equation}
where
\begin{equation}\label{JA-direct-sum-slicing}
    \mathcal{J}_{A}=\Upsilon_{A}+\bar{\nabla}_A(\Omega\mathcal{P}) \,,\qquad  {\mathcal{P}}_{(s)}=
  \begin{cases} 
    \frac{1}{s}\,e^{s\mathcal{P}}=\frac1{s} \left(\frac{\eta}{\Theta^2_l}\right)^s & \text{if } s \neq 0 \\
   \mathcal{P}       & \text{if } s =0 \, . 
  \end{cases}
\end{equation}
We call $\Omega, \mathcal{P}_{(s)}, \mathcal{J}_{A}$, respectively, entropy aspect, expansion aspect and angular momentum  aspect. The expressions above make manifest that the flux proportional to the traceless news tensor $N_{AB}$ is not integrable. Therefore, this slicing is in the family of genuine slicings.

The Raychaudhuri and Damour equations can be recast in terms of the charges
\begin{subequations}\label{field-eq-Ps-J}
    \begin{align}
        &{\mathcal{D}_{v}\mathcal{P}_{(s)}-e^{s\mathcal{P}}\Big(\Gamma+\frac{2N_{AB}N^{AB}}{\Theta_l}\Big)\approx 0}\\
        &\mathcal{D}_{v}\mathcal{J}_{A}+2\Omega\bar{\nabla}^{B} N_{AB} -2 \Omega \bar{\nabla}_A (\Theta_l^{-1} N_{BC}N^{BC})\approx 0\, .
    \end{align}
\end{subequations}
Moreover, the charges transform as
\begin{subequations}
    \begin{align}
         &\delta_{\xi}\Omega=\tilde{T}^{(s)}\, e^{s\mathcal{P}}\\
         &\delta_{\xi}\mathcal{P}_{(s)}\approx -( \delta_{s,0}+ s {\mathcal{P}}_{(s)}) \, \tilde{W} +\frac{2e^{s \mathcal{P}}  T }{\Theta_l}N_{AB}N^{AB}\\
         &\delta_{\xi}\mathcal{J}_{A}\approx \mathcal{L}_{\tilde{Y}}\mathcal{J}_{A}-2 \bar{\nabla}^B (e^{-s \mathcal{P}}\Omega  T N_{AB})+2 \bar{\nabla}_A (e^{-s \mathcal{P}}\Omega T\Theta_l^{-1} N_{BC}N^{BC})\,.
    \end{align}
\end{subequations}
Using the MB, the charge algebra is 
\begin{subequations}\label{algebra-direct-sum-basis}
    \begin{align}
        &\{\Omega(v,x),\Omega(v,x')\}=0\\
        &\{{\mathcal{P}}_{(s)}(v,x),{\mathcal{P}}_{(s')}(v,x')\}=0\\
        &\{\Omega(v,x),{\mathcal{P}}_{(s)}(v,x')\}=16\pi G\left(s {\mathcal{P}}_{(s)}(v,x)+\delta_{s,0}\right)\delta^{D-2}\left(x-x'\right)\\
        &\{\mathcal{J}_A(v,x),\mathcal{J}_B(v,x')\}=16\pi G\left(\mathcal{J}_{A}(v,x')\partial_{B}-\mathcal{J}_{B}(v,x)\partial'_{A}\right)\delta^{D-2}\left(x-x'\right)\\
        &\{\mathcal{J}_A(v,x),\Omega(v,x')\}=\{\mathcal{J}_A(v,x),{\mathcal{P}}_{(s)}(v,x')\}=0\,.
    \end{align}
\end{subequations}
This algebra is the direct sum $\mathcal{C}^{(s)}_2 \oplus$ Diff$({\cal N}_v)$, where $\mathcal{C}^{(s)}_2$ is generated by the $\Omega(v,x), {\mathcal{P}}_{(s)}(v,x)$-towers of charges and Diff$({\cal N}_v)$ by $\mathcal{J}_A(v,x)$. We call this slicing a direct-sum genuine slicing. The algebra for $s=0$ is qualitatively different from $s\neq0$. The former has a central term while all $s\neq 0$ have no central terms. For $s\neq0$, at any given point on ${\cal N}$, $\mathcal{C}^{(s)}_2$ is a two-dimensional subalgebra of $sl(2,\mathbb{R})$.\footnote{%
The case $s=-1/2$ is special as ${\cal P}_{(-1/2)}=-\frac{2\Theta_l}{\sqrt{\eta}}$ is proportional to the expansion $\Theta_l$. For $s<0$, ${\cal P}_{(s)}$ has a smooth non-expanding $\Theta_l\to 0$ limit.}

\paragraph{Heisenberg slicing.}
For $s=0$ case the charge algebra \eqref{algebra-direct-sum-basis} takes a simple form of Heisenberg $\oplus$ Diff(${\cal N}_v$). The Heisenberg slicing is in a sense a fundamental slicing, since the other genuine slicings in the $s$-family (and many others, see, e.g., \cite{Adami:2020ugu}) may be constructed from this slicing. Due to its importance as algebraic building block, we display the charges
\begin{equation}
     \tilde{Q}^{\text{I}}_{\xi}= \frac{1}{16\pi G} \int_{{\cal N}_v} \d{}^{D-2} x \left( \tilde{W} \Omega+\tilde{Y}^{A}  \mathcal{J}_{A}+\tilde{T} \mathcal{P}\right)\, ,
\end{equation}
and flux
\begin{equation}
    \tilde{F}_\xi (\delta g)= -\frac{1}{16\pi G} \int_{{\cal N}_v} \d{}^{D-2} x\  \left[ \tilde{T}-\bar{\nabla}_{C}(\Omega \tilde{Y} ^{C})\right]\Theta_l^{-1} N^{AB}\delta\Omega_{AB}
\end{equation}
 where $\tilde T=\tilde T^{(0)}$. The associated transformation laws 
\begin{subequations}
    \begin{align}
         &\delta_{\xi}\Omega=\tilde{T}\\ 
         &\delta_{\xi}\mathcal{P}\approx - \tilde{W} +\frac{2   T}{\Theta_l}N_{AB}N^{AB}\\
         &\delta_{\xi}\mathcal{J}_{A}\approx \mathcal{L}_{\tilde{Y}}\mathcal{J}_{A}-2 \bar{\nabla}^B ( \Omega  T N_{AB})+2 \bar{\nabla}_A ( \Omega T\Theta_l^{-1} N_{BC}N^{BC})
    \end{align}
\end{subequations}
yield the charge algebra
\begin{subequations}\label{Heisenberg-direct-sum-algebra}
    \begin{align}
        &\{\Omega(v,x),\Omega(v,x')\}=\{\mathcal{P}(v,x),\mathcal{P}(v,x')\}=0\\
        &\{\Omega(v,x),\mathcal{P}(v,x')\}=16\pi G\delta^{D-2}\left(x-x'\right)\\
        &\{\mathcal{J}_A(v,x),\Omega(v,x')\}=\{\mathcal{J}_A(v,x),\mathcal{P}(v,x')\}=0\\
        &\{\mathcal{J}_A(v,x),\mathcal{J}_B(v,x')\}=16\pi G\left(\mathcal{J}_{A}(v,x')\partial_{B}-\mathcal{J}_{B}(v,x)\partial'_{A}\right)\delta^{D-2}\left(x-x'\right)\,.
    \end{align}
\end{subequations}
The brackets in the first two lines above are the reason why we chose the name Heisenberg slicing.

We end this section with some additional remarks. Regardless of the slicing, we have $D$ towers of charges, which is the same number as the BDOF. Each charge is a generic function over the co-dimension one null boundary ${\cal N}$. In particular each charge is given by an integral over the transverse space ${\cal N}_v$ and therefore it has $v$ dependence. The bulk degrees of freedom are encoded in $N_{AB}, \hat{g}^{(v)}_{{AB}}(r, x^A)$ modes (see the discussion in section \ref{sec:solution space}). The latter do not enter in the charge analysis. By contrast, the news $N_{AB}$ appears in the flux. This provided the very rationale to call it news. Its transformation in the thermodynamic slicing  \eqref{N-AB-transf} is homogeneous, $\delta_\xi N_{AB}=0$ when $N_{AB}=0$. While this statement is slicing-independent, the explicit expression for $\delta_\xi N_{AB}$ is, in the  Heisenberg slicing, 
    \begin{equation}
        \delta_{\xi}N_{AB}=\mathcal{D}_{v}\left[\left(\tilde{T}-\bar{\nabla}_{A}(\Omega\tilde{Y}^{A})\right)\frac{N_{AB}}{\Omega\Theta_l}\right]+\mathcal{L}_{\tilde{Y}}N_{AB}\,.
           \end{equation} 
Having a homogeneous transformation means that action of boundary charges will not take one out of the vanishing genuine flux sector. 

We shall make further comments on slicings in the concluding section, but for now move on to another physically relevant aspect of non-integrable surface charges, the flux balance equations.

\section{Null surface balance equation}\label{sec:charge conservation balance equation}

In the presence of flux, surface charges are not integrable \cite{Wald:1999wa,Barnich:2011mi}. Moreover, non-integrability and presence of flux are closely related to the charge  non-conservation. While integrability is slicing-dependent, as discussed, there are genuine slicings for which  the flux is proportional to the genuine news $N_{AB}$ associated with infalling gravitons. Conservation, too, depends on the choice of phase space slicing. In some earlier works \cite{Adami:2020amw, Adami:2020ugu, Adami:2021sko} we have discussed the relation between  charge integrability and conservation is captured by the generalized conservation equation, which in the more standard null infinity analyses  is called ``flux balance equation'' \cite{Compere:2019gft,Godazgar:2018vmm}. In this section, we briefly discuss the null surface balance equation for the  thermodynamic and Heisenberg slicings discussed in the previous section.

\subsection{Balance equation in thermodynamic slicing}\label{sec:balance-eq-thermodynamic}

{For the thermodynamic slicing in section \ref{sec:thermoslicing},} the generator of translations along the advanced time $\partial_v$ is among the symmetry generators
$\partial_v=\xi(T=1,W=0,Y^A=0)$. {The associated integrable part of the charge}  \eqref{Integrable-part-Charge-Geometric} and {the flux} \eqref{Non-Integrable-part-Charge-Geometric} 
\begin{subequations}\label{partial-v-charge-flux}
\begin{equation}
    \mathbf{H}_v:=Q^{\text{I}}_{\partial_v} {= \frac{1}{16\pi G} \int_{{\cal N}_v}\! \d{}^{D-2} x \ \left(-\Gamma \Omega  +\mathcal{U}^{A}{\Upsilon}_A\right)}
\end{equation} 
\begin{equation}
    F_{\partial_v}(\delta g ; g) =   \frac{1}{16\pi G} \int_{{\cal N}_v}\!\! \d{}^{D-2} x \,  \Big(  -2\Omega \delta \Theta_l + \Omega \Theta_l \frac{\delta \eta}{\eta}+\Omega \delta \Gamma -\Upsilon_{A}\delta \mathcal{U}^{A} -\Omega N^{AB} \delta\Omega_{AB} \Big),
\end{equation}
\end{subequations}
{obey the null surface energy balance  equation}
\begin{equation}\label{GCCE-Hv-Geometric-slicing}
    \frac{\d{}}{\d v }
      \mathbf{H}_v \approx  -F_{\partial_v}(\delta_{\partial_v}g)
\end{equation}
where $\approx$ denotes on-shell equality and $F_{\partial_v}(\delta_{\partial_v}g):=F_{\partial_v}(\delta_\xi g ; g)|_{\xi=\partial_v}$. This flux receives two contributions, one from the bulk modes, the $N_{AB}N^{AB}$ term in $F$, and the other from boundary modes. The latter is essentially a reflection of the fact that in the thermodynamic slicing, the coordinate system adopted \eqref{G-F-M-01} corresponds to a non-inertial frame for the boundary dynamics. As viewed by the observer adopting the coordinate system $v,r, x^A$, the quantity $\mathbf{H}_v=\mathbf{H}_v (v)$ is the boundary Hamiltonian. Thus, a suggestive interpretation of \eqref{GCCE-Hv-Geometric-slicing} is that it describes an open system, the Hamiltonian of which is time-dependent as a consequence of leakage. Equation \eqref{GCCE-Hv-Geometric-slicing} is an instance of a null surface balance equation.

Similarly, one may study the time variation of all other charges, in particular of the zero mode charges, angular momentum, associated with the symmetry generator $\partial_A=\xi(T=0,W=0,Y^A=1)$, 
\begin{subequations}\label{S-JA}
\begin{equation}
    \mathbf{J}_A:=Q^{\text{I}}_{\partial_A} = \frac{1}{16\pi G} \int_{{\cal N}_v} \d{}^{D-2} x \ \Upsilon_A\qquad \qquad F_{\partial_A}(\delta g)=0
\end{equation}
and entropy, associated with the symmetry generator  $-r\partial_r=\xi(T=0,W=1,Y^A=0)$,
\begin{equation}
    \mathbf{S}:=4\pi Q^{\text{I}}_{-r\partial_r} = \frac{1}{4 G} \int_{{\cal N}_v} \d{}^{D-2} x\  \Omega\qquad\qquad  F_{-r\partial_r}(\delta g)=0\,.
\end{equation}
\end{subequations}
Both obey null surface balance equations 
\begin{subequations}\label{S-JA-time-derivative}
\begin{align}
  \frac{\d{}}{\d v }
      \mathbf{J}_A&= \frac{1}{16\pi G} \int_{{\cal N}_v} \d{}^{D-2} x\ \partial_v{\Upsilon}_A
 \approx   -F_{\partial_v}(\delta_{\partial_{A}}g)\label{eq:nolabel}\\
  \frac{\d{}}{\d v }\mathbf{S}&
      = \frac{1}{4 G} \int_{{\cal N}_v} \d{}^{D-2} x\ \Omega \Theta_l\approx  -4\pi F_{\partial_v}(\delta_{-r\partial_{r}}g)\label{S-time-derivative}\,.
\end{align}\end{subequations}
The null surface balance equation for entropy \eqref{S-time-derivative} shows that the time derivative of the area is given by the integral of the expansion, but does not involve any bulk graviton flux.  The time derivative of the angular momentum \eqref{eq:nolabel} has a term proportional to the total angular momentum of the graviton flux through the null surface and some additional terms. The latter appear because we are in a non-inertial rotating frame. 

The algebraic relations \eqref{Thermodynamic-slicing-algebra} imply
\begin{equation}\label{H-Q-xi}
    \{ \mathbf{H}_v, Q^{\text{I}}_{\xi}\}_{{\tiny{\text{MB}}}}= Q^{\text{I}}_{\partial_v\xi}\qquad \qquad  \{ \mathbf{S}, Q^{\text{I}}_{\xi}\}_{{\tiny{\text{MB}}}}=0
\end{equation}
and in particular
\begin{equation}\label{H-S-JA-bracket}
    \{ \mathbf{H}_v, \mathbf{S}\}_{{\text{\tiny{MB}}}}=\{ \mathbf{H}_v, \mathbf{J}_A\}_{{\text{\tiny{MB}}}}=  \{ \mathbf{S}, \mathbf{J}_A\}_{{\text{\tiny{MB}}}}= 0\,. 
\end{equation}
As expected, $\mathbf{H}_v$ generates time translations. Moreover, the entropy $\mathbf{S}$ commutes with all the charges. The zero mode  charges $\mathbf{H}_v, \mathbf{S}, \mathbf{J}_A$ mutually commute.

On can show that balance equations for zero-mode charges \eqref{GCCE-Hv-Geometric-slicing} and \eqref{S-JA-time-derivative},  can be generalized to all null boundary charges for generic symmetry generator $\xi$ as, 
\begin{equation}\label{Balance-Eq}
      \frac{\d{}}{\d v }Q^{\text{I}}_{\xi} = \delta_{\partial_v}Q^{\text{I}}_{\xi}+Q^{\text{I}}_{\partial_{v}\xi}\approx 
     - F_{\partial_v}(\delta_{\xi}g)
\end{equation}
by virtue of \eqref{H-Q-xi}, where we used the definition of the MB \eqref{BT-Bracket} and that $F_{\partial_v}(\delta_{\xi}g)$ is given by $F_{\partial_v}(\delta g, g)$ in \eqref{partial-v-charge-flux} evaluated at $\delta_\xi g$. 

To derive \eqref{Balance-Eq} we have used {the fact} that $\partial_v$ is among our field independent symmetry generators in the thermodynamic slicing. The null surface balance equation \eqref{partial-v-charge-flux} shows that the flux $F_{\partial_v}(\delta_{\xi}g)$ receives contributions from the genuine flux, the term proportional to $N^{AB}$, as well as from terms only involving boundary fields, referred to as fake flux. Like for the angular momentum, the latter is generically there because the $v, x^A$ coordinates do not correspond to an inertial observer at the boundary.

\subsection{Balance equation in Heisenberg slicing} \label{sec:balance-eq-Heisenberg} 

Unlike the thermodynamic slicing \eqref{H-S-JA-bracket}, the zero mode charges in the Heisenberg slicing
\begin{subequations}\label{S-JA-sfamily}
\begin{align}
    &\tilde{\mathbf{H}}:=\tilde{Q}^{\text{I}}_{\tilde{T}=1} = \frac{1}{16\pi G} \int_{{\cal N}_v} \d{}^{D-2} x \ \mathcal{P}\\ 
    &\tilde{\mathbf{S}}:=4\pi\tilde{Q}^{\text{I}}_{{\tilde{W}=1}} = \frac{1}{4 G} \int_{{\cal N}_v} \d{}^{D-2} x\  \Omega\\ 
    &\tilde{\mathbf{J}}_A:=\tilde Q^{\text{I}}_{\tilde{Y}^A=1} = \frac{1}{16\pi G} \int_{{\cal N}_v} \d{}^{D-2} x \ \mathcal{J}_A
\end{align}
\end{subequations} 
do not commute with each other. Nor does the entropy {generically} commute with the remaining charges,
\be
\{\tilde{\mathbf{S}},  Q^{\text{I}}_{\tilde{\xi}}\}= \frac{1}{4G} \int_{{\cal N}_v} \d{}^{D-2} x\ \tilde T  
\ee
implying
\be
\{\tilde{\mathbf{S}}, \tilde{\mathbf{H}}\}= \frac{1}{4G} \int_{{\cal N}_v} \d{}^{D-2} x\qquad\qquad \{\tilde{\mathbf{S}}, \tilde{\mathbf{J}}_{A}\}=\{\tilde{\mathbf{H}},\tilde{\mathbf{J}}_{A}\}=0\,.
\ee
Notably, $\tilde{\mathbf{S}}$ and $\tilde{\mathbf{H}}$ are Heisenberg pairs with an effective $\hbar$ proportional to $1/G$. One can therefore change the entropy of the system by injecting  $\tilde{\mathbf{H}}$ charge. Recall that $\tilde{\mathbf{H}}$ is the charge associated with the symmetry generator $\tilde{W}=0=\tilde Y^A$ and $\tilde T=\Omega\Theta_lT=1$, but not with unit $v$-translations, so we do not refer to it as energy. Moreover, there are no other local combinations of charges playing this role. Thus, in the Heisenberg slicing the zero-mode charge $\tilde{\mathbf{H}}$ should not be viewed as a Hamiltonian, but rather as the Heisenberg conjugate of the entropy.

Since $\partial_v$ is not among the symmetry generators in the Heisenberg slicing, we do not have a null surface balance equation like  in thermodynamic slicing \eqref{Balance-Eq}. The {zero-mode} charge dynamics is given by\footnote{%
The middle equation \eqref{Raych-Heisenberg} may also be written as $\mathcal{D}_{v}\ln \left({\Theta_l}{\Omega^{\frac{1}{D-2}}}\right)=\kappa+{\Theta_l}^{-1}{N_{AB}N^{AB}}$.}
\begin{subequations}\label{charge-EoM-Heisenberg-genuine}
    \begin{align}
         \mathcal{D}_{v}\Omega &= \Omega \Theta_l\\
        \mathcal{D}_{v}\mathcal{P} &= \Gamma
        +\frac{2N_{AB}N^{AB}}{\Theta_l}\label{Raych-Heisenberg}\\
        \mathcal{D}_{v}\mathcal{J}_{A} &= 2 \Omega \bar{\nabla}_A (\Theta_l^{-1} N_{BC}N^{BC})-2\Omega\bar{\nabla}^{B} N_{AB} \label{Damour-Heisenberg}\,.
    \end{align}
\end{subequations}

\section{Vanishing genuine news}\label{sec:Integrable-cases}

An interesting special case arises when the news $N_{AB}$ vanishes, which is the focus of this section. Generically, the expansion does not have to vanish, $\Theta_l\neq 0$. However, if vanishing expansion is assumed, $\Theta_l=0$, then vanishing news is implied as consequences of the Raychaudhuri equation \eqref{EoM-Raychaudhuri}. The main goal of this section is to exhibit the subtle differences between the generic situation, $N_{AB}=0\neq \Theta_l$, and vanishing expansion, $N_{AB}=0=\Theta_l$.

\subsection{Generic situation}\label{sec:Vanishing-Flux}

Assuming $N_{AB}=0$, several of our previous results simplify, like the Raychaudhuri and Damour equations \eqref{field-eq-Ps-J} 
\begin{equation}
    \mathcal{D}_{v}\mathcal{P}= \Gamma \qquad  \mathcal{D}_{v}\mathcal{J}_{A}= 0\,.
\end{equation}
There exists a co-rotating frame where the angular momentum aspect $\mathcal{J}_A$ is $v$ independent, $\mathcal{J}_A=\mathcal{J}_A(x^B)$.\footnote{%
The entropy aspect $\Omega$ and the expansion aspect ${\cal P}$ still depend on $v$ in this co-rotating frame. Alternatively, one can find a co-expanding frame, through the choice of normalization of the vector normal to the null surface and appropriate adjustment of the non-affinity parameter $\kappa$, such that ${\cal D}_v \Theta_l=0$, or a frame in which ${\cal D}_v {\cal P}=0$. In this frame ${\cal P}$ can be made $v$-independent. 
}

For the analysis of charges, one needs to choose a slicing. Let us start with the direct-sum genuine slicings introduced in section \ref{sec:genuine-slicing} for which the charges 
\begin{equation}
     \delta {Q}_{\xi}= \frac{1}{16\pi G} \int_{{\cal N}_v} \d{}^{D-2} x \left( \tilde{W} \delta\Omega+\tilde{Y}^{A}  \delta\mathcal{J}_{A}+\tilde{T}^{(s)} \delta\mathcal{P}_{(s)}\right),
\end{equation}
are integrable and obey the algebra  \eqref{algebra-direct-sum-basis}. 

In non-genuine slicings the situation is more complicated, in general, due to fake news. Studying particularly the thermodynamics slicing would be a direct extension of the analysis of \cite{Adami:2020amw} to $D>4$. Since the physical discussion is going to be very similar to the one in \cite{Adami:2020amw}, we refer the reader to that work instead of displaying these results.

\subsection{Vanishing expansion}\label{sec:non-expanding}

For non-expanding null boundaries, $\Theta_l=0$, the Raychaudhuri equation \eqref{EoM-Raychaudhuri} enforces vanishing news, $N_{AB}=0$. We address now three different slicings to highlight some new features as compared to the generic situation, $\Theta_l\neq 0$.

\paragraph{Thermodynamic slicing.} A careful analysis of the charges reveals that $T$ {generates trivial diffeomorphisms, so we have one tower of charges less.} One may use this fact to gauge fix $\eta=1$, see section 6 of \cite{Adami:2020amw} for a similar, but more detailed analysis. Therefore, the boundary phase space in this case is labeled by $\Omega$ and ${\Upsilon}_A$, only. See \cite{Adami:2021kvx} for a more detailed discussion. 

\paragraph{{Direct-sum} genuine slicing.} The transformation to the genuine slicing \eqref{tilde-slicing-generic-s} and also the tower of ${\cal P}$ charges \eqref{cal-P--def} are ill-defined for $\Theta_l=0$. Revisiting the analysis shows that the charge associated with $\tilde{T}$ vanishes. Hence, we remain with only two towers of integrable charges,
\begin{equation}\label{chargenonexp}
     {\delta} Q_{\xi}= \frac{1}{16\pi G} \int_{{\cal N}_v} \d{}^{D-2} x \left(\tilde{W}\delta\Omega+\tilde{Y}^{A}\delta\Upsilon_{A}\right)
\end{equation}
where $\tilde W=W-\Gamma \, T$ and $\tilde Y^A=Y^A+{\cal U}^A\,T$. The Damour equation
\begin{equation}
    \mathcal{D}_{v}\Upsilon_{A}+\bar{\nabla}_A(\Omega\Gamma)=0
\end{equation}
fixes the $v$-dependence of $\Upsilon_A$ in terms of $\Omega$, $\Gamma$, ${\cal U}^A$.  Moreover, $\Theta_l=0$ implies ${\cal D}_v \Omega=0$ and therefore the $v$-dependence of $\Omega$ is also fixed in terms of ${\cal U}^A$. Note, however, that the $v$-dependences of $\Gamma$ and ${\cal U}^A$ are still arbitrary. So, in general our charges $\Omega, \Upsilon_A$ depend arbitrarily on $v$  through $\Gamma, {\cal U}^A$.

The charge transformation laws 
\begin{equation}
        \delta_\xi\Omega=\Omega \bar{\nabla}_A\tilde{Y}^{A} \qquad\qquad
        \delta_{\xi}\Upsilon_{A}\approx \mathcal{L}_{\tilde{Y}}\Upsilon_{A}+\Omega\bar{\nabla}_A {\tilde{W}}
\end{equation}
yield the charge algebra
\begin{subequations}\label{BMS-like-non-expanding-algebra}
    \begin{align}
        \{\Omega(v,x),\Omega(v,x')\}&=0\\
        \{\Upsilon_{A}(v,x),\Omega(v,x')\}&=-16\pi G\Omega(v,x)\partial_{A}'\delta^{D-2}(x-x')\\
        \{\Upsilon_{A}(v,x),\Upsilon_{B}(v,x')\}&=16\pi G\left(\Upsilon_{A}(v,x')\partial_{B}-\Upsilon_{B}(v,x)\partial_{A}'\right)\delta^{D-2}(x-x')\,.
    \end{align}
\end{subequations}
The algebra above is isomorphic to the near horizon symmetry algebra in one of the slicings introduced in \cite{Grumiller:2019fmp} with $s=1$. This is not surprising, since vanishing expansion was built into the boundary conditions enforced in that work.

\paragraph{Heisenberg-like slicing.} Upon the change of slicing
\begin{equation}
    {\cal J}_{A}^{\text{\tiny{H}}}=\Omega^{-1}\Upsilon_{A}
    \qquad\qquad
Y^A_{{\text{H}}}=\Omega \hat{Y}^A\qquad\qquad    W_{{\text{H}}}=\hat W+ \Omega^{-1} \hat{Y}^A {\cal J}_A
\end{equation}
the algebra above simplifies further,
\begin{subequations}\label{Heisenberg-like-non-expanding}
    \begin{align}
        \{\Omega(v,x),\Omega(v,x')\}&=0\\
        \{ {\cal J}_{A}^{\text{\tiny{H}}}(v,x),\Omega(v,x')\}&=16\pi G \partial_{A}\delta^{D-2}(x-x')\\
        \{ {\cal J}_{A}^{\text{\tiny{H}}}(v,x), {\cal J}_{B}^{\text{\tiny{H}}}(v,x')\}&=16\pi G \Omega^{-1}(v,x)\mathcal{F}_{BA}\delta^{D-2}(x-x')
    \end{align}
\end{subequations}
where $\mathcal{F}_{AB}:=\partial_{A} {\cal J}_{B}^{\text{\tiny{H}}}(x)-\partial_{B}' {\cal J}_{A}^{\text{\tiny{H}}}(x')$. This algebra is the same as the Heisenberg-like algebra of \cite{Grumiller:2019fmp}, where our charge $\Omega$ is equivalent to their charge $\mathcal{P}$. Again our charges can depend on $v$.

Note that the Heisenberg-like algebra \eqref{Heisenberg-like-non-expanding} differs from the Heisenberg algebra discussed in \eqref{Heisenberg-direct-sum-algebra}. In particular, here we do not have the expansion aspect ${\cal P}$ among our generators and the Heisenberg conjugate of $\Omega$ is now the exact part of angular momentum aspect ${\cal J}_A$ (see \cite{Grumiller:2019fmp} for more discussion). Another important difference to that work is that the entropy, the zero mode charge proportional to the integral of $\Omega$, does not generically commute with the other charges, though it does commute at least with the zero mode charge of the angular momentum aspect. 

As summary, we contrast the generic situation for vanishing news with the special case of vanishing expansion. Generically, we obtained three towers of integrable charges for genuine slicings. In the non-expanding case we lost one charge tower, as a consequence of the Raychaudhuri equation. Technically, this is so because the absence of expansion renders $\eta$ pure gauge, and the boundary phase space therefore has one less function in it. 

In conclusion, when considering vanishing news it is crucial to additionally specify whether or not expansion also is assumed to vanish, since the associated boundary phase spaces have different dimensions, depending on this choice.

\section{Null boundary memory effects} \label{sec:GW-through-horizon}

In this section, we apply our charge and flux analysis to a physically interesting example. Suppose that a gravitational shockwave passes through the horizon of a  black hole and the system at late times again settles into another  black hole. We expect the information about the gravitational wave to be encoded in changes in the surface charges. This physical process is depicted in Fig.~\ref{fig:shockwave}. 

We call the persistent change of the surface charges due to the absorption of such a shockwave \textit{null boundary memory effect}, by analogy to memory effects at the celestial sphere \cite{Strominger:2014pwa, Donnay:2018ckb, Compere:2018ylh, Bhattacharjee:2020vfb,Rahman:2019bmk}. Historically, imprints of gravitational waves on detectors were discovered in \cite{Zeldovich:1974gvh} and the term memory effect coined in \cite{Braginsky:1985vlg}, see also \cite{Braginsky-Thorne}. The original (displacement) memory effect is a change in the relative position of pairs of detectors after passage of some burst of gravitational waves. In the recent literature many other memory effects have been discussed that are mainly associated with asymptotic symmetries and soft gravitons, see \cite{Zhang:2017rno, Strominger:2017zoo}.

We start with the Schwarzschild black hole of horizon radius $r_h$, 
\be
\Theta_{n}=-{4}\kappa=-\frac{{2}}{r_{h}}\qquad\qquad \eta=1\qquad \qquad \Theta_l={\cal U}^A={\cal J}_A=L_{AB}=N_{AB}=0\,. 
\ee 
See appendix \ref{appen:Kerr-Metric-GNC} for a generalization to the Kerr black hole.

\def \L {5} 
\definecolor{darkgreen}{HTML}{006622}
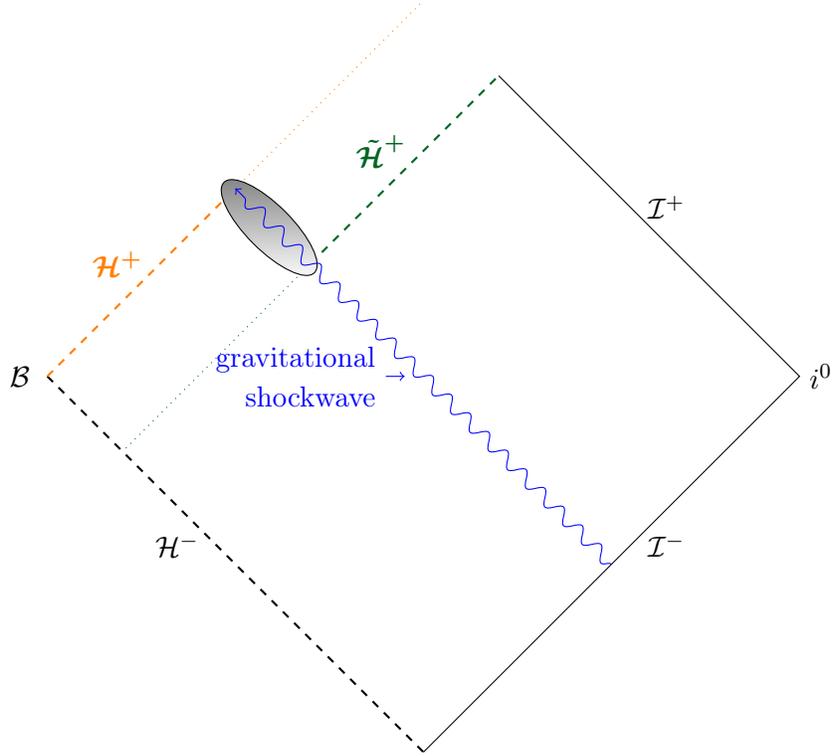
\begin{figure}
    \centering
\begin{tikzpicture}
%
  \draw[thick,black,dashed] (-\L,0) coordinate (bif) -- (0,-\L) coordinate (imi);
  \draw[thick,orange,dashed] (bif) -- (-0.5*\L,0.5*\L) coordinate (mid);
  \draw[thin,orange,dotted] (mid) -- (0.0*\L,1.0*\L) coordinate (top);
  \draw[thick,darkgreen,dashed] (-0.3*\L,0.3*\L) coordinate (mas) -- (0.2*\L,0.8*\L) coordinate (end);
  \draw[thin,darkgreen,dotted] (mas) -- (-0.8*\L,-0.2*\L);  \shadedraw[rotate=-45] (-0.57*\L,-0.01*\L) ellipse (8.5mm and 3mm);
  \draw[black] (end) -- (1.0*\L,0.0*\L) node[right] (io) {$i^0$} -- (imi);
  \draw[black] (-1.02*\L,0) node[left] (scrip) {$\mathcal{B}$}
               (0.57*\L,0.45*\L) node[right]  (scrip) {$\mathcal{I}^+$}
               (0.57*\L,-0.45*\L) node[right] (scrip) {$\mathcal{I}^-$}
               (-0.57*\L,-0.45*\L) node[left] (scrip) {$\mathcal{H}^-$}
               (-0.72*\L,0.3*\L) node[left] (scrip) {\color{orange}$\boldsymbol{\mathcal{H}^+}$}
               (-0.02*\L,0.6*\L) node[left] (scrip) {\color{darkgreen}$\boldsymbol{\mathcal{\tilde{H}}^+}$};    
  \draw[blue,snake it,->] (0.5*\L,-0.5*\L) coordinate (scr) -- (mid);
  \draw[blue,<-] (-0.05*\L,0.0*\L) -- (-0.1*\L,0.0*\L) node[left,align=right] {gravitational\\shockwave};
\end{tikzpicture}
    \caption{Penrose diagram for {\color{blue}{shockwave}} entering black hole. Shaded oval denotes absorption (not in solution space). Dashed {\color{orange}{orange}} ({\color{darkgreen}{green}}) line is initial (final) horizon $\color{orange}{\bc{H}^+}$  ($\color{darkgreen}{\tilde{\bc{H}}^+}$).}
    \label{fig:shockwave}
\end{figure}
We focus on the co-rotating  ${\cal U}^A=0$ case and consider a burst of gravitational waves that passes through the null surface around advanced time $v=v_0$; specifically, we design the news function as
\begin{equation}\label{shockwave-response}
    N_{AB}=\bar{N}_{AB}\ f(v-v_0) 
\end{equation}
where $\bar{N}_{AB}$ is a dimensionless symmetric traceless tensor on $\mathcal{N}_v$. The profile function $f(v-v_0)$ specifies the time dependence of the incident shockwave, and we choose it to be of delta-function type, sharply peaked around $v=v_0$.
\be\label{delta-reguliarized}
f(v-v_0)= 
\sqrt{\frac{2}{\pi}} \frac{\epsilon^{3/2}}{\epsilon^2+(v-v_0)^2}\qquad\qquad 0<\epsilon\ll 1
\ee
The normalization of $f$ is chosen such that $\int_{-\infty}^\infty\extd v\, f^2=1$.

Initially and finally, the system is stationary by assumption and has vanishing expansion, $\Theta_l=0$. As a response to the incident wave, the expansion is non-zero for a short period. In the initial and final stages the system is described by two towers of boundary charges $\Omega, {\cal J}_A$, as discussed in section \ref{sec:non-expanding}. Specifically, the system is assumed to satisfy the early- and late-times conditions ($|v-v_0|\gg \epsilon$)
\be\label{RD-boundary-cond}
\Omega=\Omega^\pm\qquad\qquad \Gamma=-2\kappa^\pm\qquad\qquad \Omega_{AB}=\bar\Omega_{AB}^\pm\qquad\qquad \Theta_l=\partial_v {\cal J}_A=0
\ee
where $\Omega^\pm, \kappa^\pm$ and $\bar\Omega_{AB}$, respectively, denote area density, surface gravity, and  metric on ${\cal N}_v$, before ($-$) and after ($+$) the passage of the wave (see again Fig.~\ref{fig:shockwave}). The area theorem (see e.g.~\cite{Chrusciel:2000cu}) implies $\Omega^+>\Omega^-$.

At early and late times, the system is described by two towers of charges, whereas during the encounter time $|v-v_0|={\cal O}(\epsilon)$ all three towers of charges, including ${\cal P}$, {can} take non-zero values. Since initially $\Theta_n\neq 0$, the $\Theta_n \Theta_l$-term in \eqref{Thetan-EoM} gives  non-trivial dynamics to $\Theta_n$. Similarly, the $\Theta_n N_{AB}$ term in \eqref{trace-less-AB} is a source for $L_{AB}$. Therefore, all modes are eventually turned on due to the passage of the gravitational wave. 
We do not solve these equations here, but merely use them to extract memories imprinted in the boundary charges after the system settled down in its new stationary point.

\subsection{Null surface expansion memory effect}\label{sec:Expansion-memory}

To specify the $v$-dependence of $\Omega$, we take a closer look at the Raychaudhuri equation \eqref{EoM-Raychaudhuri},
\be\label{Raych-N}
\partial_v\Theta_l-\kappa \Theta_l+\frac{\Theta_l^2}{D-2} +N_{AB} N^{AB}=0\qquad\qquad \Theta_l=\partial_v\ln\Omega
\ee
with boundary conditions \eqref{RD-boundary-cond}. The equation \eqref{Raych-N} differs from the usual focusing equation by the term $\kappa\Theta_l$. For early and late times, $|v-v_0|\gg \epsilon$, the $N^2$ term drops out and \eqref{Raych-N} has two fixed points, $\Theta_l=0$ and $\Theta_l=(D-2)\kappa$. For $\kappa>0$, $\Theta_l=0$ is a repulsor and $\Theta_l=(D-2)\kappa$ an attractor. Therefore, the system cannot settle in a stationary black hole of vanishing $\Theta_l$ and our desired boundary conditions \eqref{RD-boundary-cond} cannot be satisfied. 

This apparent inconsistency could be resolved as follows. During the absorption process the locus $r=0$ does not remain a null surface, so our setup in the present work is insufficient to describe it. Inevitably, we need to consider another mode, switched off by our assumptions in section \ref{sec:metric-expansion-prelim}, namely an ${\cal O}(1)$ term in $V$ in the expansion \eqref{nearN-expansion}, which relaxes the condition that our boundary $\cal N$ is null. See \cite{Geiller:2021vpg} for the $D=3$ example. This generalization adds an extra freedom and a corresponding new charge. So, to fully follow the dynamics of the absorption process one should use the generalized form of Raychaudhuri equation given in \eqref{eom-ll}, where the last term in that equation can resolve the inconsistency discussed above. A full analysis of the absorption process is beyond the scope of this work.

Instead, we simply assume that the inconsistency can be resolved along the lines above and study a $v$-integrated version of \eqref{Raych-N} to extract a memory effect. We treat the incident gravitational wave as a perturbation of the existing black hole and keep terms up to ${\cal O}(N^2)$. Multiply \eqref{Raych-N} by $\Omega$ and integrate over $v$. The term $\Theta_l^2$ is negligible since it is suppressed as compared to the linear terms in $\Theta_l$. The integrated term coming from $\partial_v\Theta_l$ is subleading as well, which can be shown as follows. While $\Omega={\cal O}(1)$, its first derivative is subleading, $\partial_v\Omega={\cal O}(N^2)$. The expression $\int\Omega\partial_v\Theta_l=\int\partial_v^2\Omega-\int\frac1\Omega\,(\partial_v\Omega)^2$ has a first term that integrates to zero and a second term of order ${\cal O}(N^4)$. The only two remaining terms, both of order ${\cal O}(N^2)$, integrate to the relation
\be\label{1st-law}
\Delta \Omega = \int\limits_{-\infty}^\infty\extd v\, \frac{\Omega}{\kappa} \,N_{AB} N^{AB}\,.
\ee
At early and late times, we expect $\kappa$ to be a constant. For a Schwarzschild black hole of mass $M$, $\kappa\sim 1/M$,  the change in $\kappa$ during the process is expected to be $\sim \Delta M/M^2$, where $\Delta M$ is proportional to $N^2$. Therefore, effects from the $v$-dependence of $\kappa$ are expected to be subleading in $N^2$ so that to a good approximation $\kappa$ is constant in $v$ and can be taken out of the integral, which we shall always do below. The result \eqref{1st-law} captures a null surface memory effect, describing how the volume form $\Omega$ changes from early to late times, $\Delta\Omega=\lim_{v\to\infty}\Omega-\lim_{v\to-\infty}\Omega$, depending on the news $N_{AB}$ associated with the gravitational shockwave. We refer to it as null surface expansion memory effect. Since the integrand in \eqref{1st-law} is non-negative (for positive $\kappa$), also $\Delta\Omega$ is non-negative, in accordance with the area theorem.

This memory effect can be rephrased suggestively as
\be\label{1st-law-memory}
T\, \Delta \bc{S} = \frac{1}{8\pi G} \int\limits_{-\infty}^\infty\extd v\, \Omega\, N_{AB} N^{AB}=:\bc{E}_{\text{\tiny{GW}}}
\ee
where $T=\frac{\kappa}{2\pi}$ is the temperature, $\bc{S}=\frac{\Omega}{4G}$ is the entropy aspect, and $\bc{E}_{\text{\tiny{GW}}}$ is the total energy density carried by the gravitational wave through ${\cal N}_v$. The above equation is a spatially local and temporally non-local energy conservation equation on ${\cal N}_v$, in contrast to usual expressions for gravitational wave energy (see e.g.~\cite{Will:1996zj}) which are spatially non-local and temporally local.

The null surface expansion memory effect \eqref{1st-law-memory} 
shows how the boundary degrees of freedom respond to the passage of the gravitational shockwave. It relates the change in the entropy aspect $\bc{S}$ to the energy passed through the surface. Unlike the memory effects discussed in the recent literature, see e.g.~\cite{Strominger:2017zoo}, this memory effect involves gravitational waves that are not soft.

\subsection{Null surface spin memory effect} \label{sec:spin-memory}

In a similar way one can work out a spin memory effect. Variation of the angular momentum charge due to passage of the shockwave may be computed integrating \eqref{Damour-Heisenberg} over $v$,
\be\label{spin-memory-effect}
        \Delta\mathcal{J}_{A}=\int\limits_{-\infty}^\infty\d v \, 2\Omega\big(\bar{\nabla}_A(\Theta_l^{-1}N_{CD}N^{CD})-\bar{\nabla}^{B}N_{AB}\big)\, .
\ee
The spin memory effect \eqref{spin-memory-effect} relates the change in black hole angular momentum aspect, $\Delta\mathcal{J}_A=\lim_{v\to\infty}\mathcal{J}_A-\lim_{v\to-\infty}\mathcal{J}_A$, to variations of the news function $N_{AB}$ along the transverse directions.

A precise evaluation of the integrand in \eqref{spin-memory-effect} requires again the extension of our analysis that we addressed already, i.e., to use \eqref{eom-lA} instead of \eqref{Damour-Heisenberg}. It is again possible to work perturbatively in the news by analogy to the previous section; however we do not present details of such an analysis here. Perturbatively, the dominant contribution to the spin-memory effect comes from the second term (linear in the news $N$). For $N_{AB}$ given in \eqref{shockwave-response}, \eqref{delta-reguliarized}, the null surface spin memory, $\Delta\mathcal{J}_A \simeq 2\sqrt{\pi\epsilon/2}\, \Omega \, \bar\nabla^B \bar N_{AB}$, vanishes in the limit $\epsilon\to 0$, unless the $\sqrt{\epsilon}$ factor is compensated by strong spatial gradients from the $\bar\nabla^B$ derivative of $\bar N_{AB}$.

\section{Discussion and concluding remarks} \label{sec:discussion}

We constructed a complete solution space for $D$-dimensional Einstein gravity in presence of a given null surface ${\cal N}$. We studied null boundary symmetries and associated $D$ towers of charges that are functions over ${\cal N}$. This work generalizes our earlier work \cite{Adami:2020amw} in three ways: (1)~It is for generic dimension $D$; (2)~we included $v$-dependence in the Diff(${\cal N}_v)$ sector of the symmetry algebra, and (3)~we discussed various different slicings of the solution space, in particular genuine slicings in which the charges become integrable in the absence of genuine news. As in other examples \cite{Grumiller:2019fmp, Adami:2020ugu, Adami:2020amw, Ruzziconi:2020wrb, Adami:2021sko, Geiller:2021vpg, Grumiller:2021cwg}, the algebra of the integrable part of the charges does depend on the slicing. In particular, there exists a Heisenberg slicing where the symmetry algebra is Heisenberg~$\oplus$~Diff(${\cal N}_v$), where ${\cal N}_v$ is the transverse surface, i.e., a co-dimension two spacelike section on ${\cal N}$. 

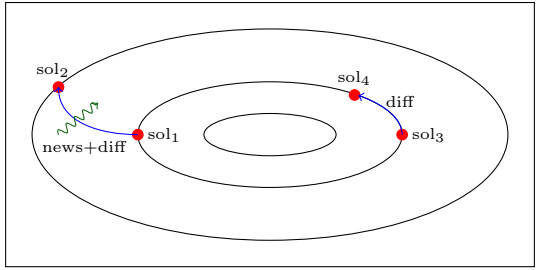
\begin{figure}
    \centering
\begin{tikzpicture}
\draw (0,0) ellipse (25pt and 8pt);
\draw (0,0) ellipse (50pt and 20pt);
\draw (0,0) ellipse (90pt and 40pt);
\draw (-100pt,-50pt) rectangle (100pt,50pt);
\filldraw [red] (50pt,0pt) circle (2pt);
\filldraw [red] (32pt,15pt) circle (2pt);
\draw[blue,->] (50pt,0pt) arc (0:48:50pt and 20pt);
\node[right=] at (40pt,12.5pt) {\tiny{diff}};
\node[right=] at (50pt,0pt) {\tiny{sol$_{3}$}};
\node[above=] at (32pt,15pt) {\tiny{sol$_{4}$}};
\filldraw [red] (-50pt,0pt) circle (2pt);
\filldraw [red] (-80pt,18pt) circle (2pt);
\draw [blue,->] (-50pt,0pt)  to [out=180,in=270] (-80pt,18pt);
\draw[green,->, snake=coil,segment amplitude=1.5pt,segment aspect=0,segment length=4pt] (-80pt,0pt)--(-65pt,12pt) ;
\node[right=] at  (-50pt,0pt) {\tiny{sol$_{1}$}};
\node[above=] at  (-82pt,18pt) {\tiny{sol$_{2}$}};
\node[right=] at (-90pt,-5pt) {\tiny{news$+$diff}};
\end{tikzpicture}
\caption{Solution space, schematically. Each point represents a solution, labeled by surface charges. On the right, a non-trivial diffeomorphism moves along some (coadjoint) orbit of the symmetry algebra from solution sol$_3$ to solution sol$_4$. On the left, additionally genuine or fake fluxes are switched on, moving from one orbit (sol$_1$) to another (sol$_2$).
}\label{fig:orbits-1}
\end{figure}

The organization of states in the solution space depends on the slicing.  Once the slicing is specified, a configuration or state is characterized by its $D$ towers of integrable charges (some of which might be zero). Configurations  in the solution space fall into coadjoint orbits of the algebra of these $D$ charges. When the boundary charges are integrable, one can label the orbits with charges associated to Killing or exact symmetries as they commute with boundary charges \cite{Hajian:2015xlp, Compere:2015knw}. Hence, coajdoint orbits are closed and one cannot  move from one orbit to another by the action of symmetries. However, when the charge variation is not integrable, acting with a symmetry that produces genuine or fake flux can move between the orbits. See Fig.~\ref{fig:orbits-1} for a schematic presentation.

To obtain the solution space, we left boundary conditions unspecified and also did not consider the variational principle. As a result, the dynamics of the $D-1$ boundary modes ${\kappa}, {\cal U}^A$ or associated surface charges remained unspecified. This latter can be fixed through an appropriate choice of boundary Lagrangian, which we leave for future work.

As discussed in section \ref{sec:GW-through-horizon}, the solution space considered here can  be extended by the addition of one extra mode: one can relax ${\cal N}$ to be a given null surface. This will add one symmetry generator $r\to r+\mu (v,x^A)$. See \cite{Geiller:2021vpg} for an explicit realization in three dimensions. Our preliminary analysis shows that adding this freedom would yield $D+1$ tower of charges. We plan to present a full analysis of this case in upcoming work. 

In section \ref{sec:GW-through-horizon} we established two new memory effects, associated with a null hypersurface, e.g., a black hole horizon: null surface expansion and null surface spin memory effects. These memory effects involve real gravitons and genuine news passing through a null surface rather than soft gravitons arriving at null infinity. Moreover, this analysis makes it apparent that the boundary modes are a substitute for the modes on one side of the boundary, e.g., $r<0$ region in Fig.~\ref{Fig:Null-surface-flux-Horizon}, which is cut out for an observer who has only access to $r\geq 0$ region. Conceptually, this is the same idea put forward in the membrane paradigm \cite{Thorne:1986iy, Price:1986yy, Parikh:1998mg}, but we formulated it through boundary degrees of freedom and surface charges as outlined in \cite{Grumiller:2018scv}. This viewpoint deserves to be explored further. 

Other interesting generalizations for future work are the inclusion of matter degrees of freedom and to investigate their interplay with boundary conditions, charges and fluxes.

\section*{Acknowledgement}
We thank Sajad Aghapour, Laura Donnay, Laurent Freidel and  Mohammad  Vahidinia for discussions. DG was supported by the Austrian Science Fund (FWF), projects P~30822, P~32581 and P~33789. MMShJ would like to acknowledge SarAmadan grant No. ISEF/M/400122. The work of VT is partially supported by IPM funds. 
CZ was supported by the FWF, projects P~30822 and M~2665. Research at Perimeter Institute is supported in part by the Government of Canada through the Department of Innovation, Science and Economic Development Canada and by the Province of Ontario through the Ministry of Colleges and Universities.

\appendix

\newcommand{\NHS}{null hypersurface symmetry}

\section{Solution space for Gaussian null-like coordinates} \label{appen:null-boundary-EOM}

In sections \ref{sec:metric-expansion-prelim} and \ref{sec:solution space} we constructed the solution space  assuming Taylor expandability of the metric around a null surface at $r=0$. In this appendix, we write the Einstein equations in Gaussian null-like gauge \eqref{G-F-M-01} without  making a Taylor expansion. We discuss solutions of these equations and show, assuming smoothness around $r=0$ of the transverse surface, that they yield the same solution space discussed in section \ref{sec:solution space}.

Consider the metric \eqref{G-F-M-01} ($\mu,\nu=\{v,r,x^A\}$, $A,B=1,\cdots, D-2$),
\begin{equation}\label{metric-appendix-B}
g_{\mu \nu}=
    \begin{pmatrix}
-\eta V +g_{AB}U^A U^B &\, \eta \,& g_{BC}U^C \\
\eta & \, 0 \, & 0 \\
 g_{AD}U^D & \, 0 \, & g_{AB}
\end{pmatrix}
\end{equation}
where we raise and lower indices by the $D-2$ dimensional metric $g^{AB}$ and $g_{AB}$, respectively. Let $\tilde{\nabla}_A$ denote the covariant derivative compatible with $g_{{AB}}$ and ${\cal G}:= \sqrt{\det g_{{AB}}}$, such that {$\Omega :={\cal G}(r)|_{r=0}$.}

Consider the two null vectors fields $n, l$ \eqref{gennullbndryl} and the $D-2$ dimensional projected metric $q_{AB}$ \eqref{g-q-nl}. We define two {two-}tensors
\begin{subequations}\label{B-tensor-generic-r}
    \begin{align}
       \tilde B^l _{\mu \nu} &:=  q^{\alpha}_{\mu} q^{\beta}_{\nu}\nabla_{\beta} l_{\alpha}  \\
        \tilde B^n _{\mu \nu} &:=  q^{\alpha}_{\mu} q^{\beta}_{\nu}\nabla_{\beta} n_{\alpha}  \, .
    \end{align}
\end{subequations}
and decompose them into trace, symmetric trace-less and anti-symmetric parts,
\begin{subequations}\label{B-l-n}
    \begin{align}
   \tilde     B^l_{\mu \nu} = & \frac{1}{D-2}   \theta_l\, q_{\mu \nu} + \sigma^l_{\mu \nu} + \omega^l _{\mu \nu} \\
  \tilde       B^n _{\mu \nu} = &\frac{1}{D-2}   \theta_{{n}}\, q_{\mu \nu} + \sigma^{\text{\tiny $n$}} _{\mu \nu} + \omega^{\text{\tiny $n$}} _{\mu \nu} \, .
    \end{align}
\end{subequations}
Note that \eqref{B-tensor-generic-r} are defined at arbitrary $r$, whereas the counterparts in \eqref{B-tensors-r=0} are defined at $r=0$. The twist tensors vanish, $\omega_{\mu\nu}^{\text{\tiny $l$}}=0, \, \omega_{\mu\nu}^{\text{\tiny $n$}}=0$. Therefore, $B^l_{\mu \nu} =\frac{1}{2} q^{\alpha}_{\mu} q^{\beta}_{\nu}\mathcal{L}_{l} g_{\alpha \beta}$ and $B^n_{\mu \nu}=\frac{1}{2} q^{\alpha}_{\mu} q^{\beta}_{\nu}\mathcal{L}_{n} g_{\alpha \beta}$ are, respectively, extrinsic curvatures of null surfaces generated by vector fields $l^\mu$ and $n^\mu$. Expansions are given as
\begin{equation}
{\theta_l=\frac{\tilde{\mathcal{D}}_v{{\cal G}}}{{{\cal G}}}+\frac{V}{2}\frac{\partial_{r}{{\cal G}}}{{{\cal G}}}\qquad\qquad
    \theta_{n}=-\frac{1}{\eta}\frac{\partial_{r}{{\cal G}}}{{{\cal G}}}}
\end{equation}
where $\tilde{\mathcal{D}}_v = \partial_v - \mathcal{L}_U$. The expansions $\theta_l$ and $\theta_n$ can depend on the radial coordinate $r$. For the metric coefficients in \eqref{nearN-expansion}, $\Theta_l=\theta_l(r=0)$ and $\Theta_n=\theta_n(r=0)$.

The shear tensors associated with the vector fields $l^{\mu}, n^\mu$ are
\begin{align}
    \sigma^l_{{AB}}&=\frac{1}{2} \mathcal{L}_l g_{{AB}}-\frac{\theta_l}{D-2}g_{{AB}}=\frac{1}{2}\tilde{\mathcal{D}}_v g_{{AB}}-\frac{\theta_l}{(D-2)}g_{{AB}}+ \frac{V}{4} \partial_r g_{{AB}}\\
    \sigma^l_{vv}&=U^{A}U^{B} \sigma^l_{{AB}}\\
    \sigma^l_{vA}&=U^{B} \sigma^l_{{AB}}\\
    \sigma^n_{{AB}}&=-\frac{1}{2\eta} \partial_r g_{{AB}}-\frac{\theta_{n}}{(D-2)}g_{{AB}}\\
    \sigma^n_{vv}&=U^{A}U^{B} \sigma^{^n}_{{AB}}\\
    \sigma^n_{vA}&=U^{B}\sigma^n_{{AB}}\,.
\end{align}
The components that are not displayed vanish. For completeness we also evaluate the H$\grave{\text{a}}$ji$\check{\text{c}}$ek one-form $\tilde {\mathcal{H}}_{\mu} = q_\mu{}^\nu l_\lambda \nabla_\nu n^\lambda$, 
\begin{equation}\label{H-a-def}
\tilde {\mathcal{H}}_{v}  =-{U}^A \tilde {\mathcal{H}}_{A} \qquad\qquad \tilde {\mathcal{H}}_{r}  =0 \qquad\qquad \tilde {\mathcal{H}}_{A}  = \frac{1}{2\eta}(-g_{{AB}}\partial_{r}U^{B}+\partial_{A}\eta)\,.
\end{equation}

The vacuum Einstein equations in $D$ dimension may be decomposed into to four scalar equations $\mathcal{E}_{ll}$, $\mathcal{E}_{ln}$, $\mathcal{E}_{nn},{\cal E}:=g^{AB}\mathcal{E}_{{AB}}$, two vector equations $\mathcal{E}_{lA}$, $\mathcal{E}_{nA}$, and a traceless tensor equation $\mathcal{E}_{{AB}}-\frac{1}{D-2} {\cal E} g_{AB}$. 

\paragraph{Scalar equations.} We list, respectively, $\mathcal{E}_{nn}=0, g^{AB}\mathcal{E}_{{AB}}=0, \mathcal{E}_{ll}=0, \mathcal{E}_{ln}=0$,
\begin{subequations}
\begin{align}
&
-\frac{1}{\eta} \partial_{r}\theta_{n}+\frac{\theta_{n}^2}{(D-2)}+\sigma_{n}^{2}=0,\label{eom-nn}\\
&
\frac{1}{\eta \sqrt{{\cal G}}}\partial_{r}\left(\sqrt{{\cal G}} \, \theta_l\right)+ \beta^{A}\beta_{A}+\tilde{\nabla}_{A}\beta^{A}-\frac{1}{2\sqrt{{\cal G}}}(\tilde{R}-2\Lambda)=0,\label{eom-AB-trace}\\
&
  \tilde{\mathcal{D}}_{v}\theta_l-\tilde \kappa\theta_l+\frac{\theta_l^2}{(D-2)}+\sigma_{{l}}^{2}-\eta\beta^{A}\tilde{\nabla}_{A}V-\frac{\eta}{2}\widetilde{\Box}V+\frac{V}{2}\partial_{r}\theta_l=0\label{eom-ll}\\
&
    \frac{1}{2\eta}\partial_{r}^2 V-\beta^{A}\Big(3\beta_{A}-\frac{2\tilde{\nabla}_{A}\eta}{\eta}\Big)+\frac{(D-3)}{(D-2)}\theta_l\theta_{{n}}-\sigma_{{n}}\cdot \sigma_{{l}} 
    +\frac{\tilde{R}}{2}-\frac{(D-4)}{(D-2)}\Lambda=0\label{eom-ln}
\end{align}
\end{subequations}
where  $\tilde R_{AB}$ is the Ricci tensor of $g_{AB}$, ${\widetilde{\Box}}=\tilde \nabla^2$, and
\begin{equation}
\sigma_{{l}}^2:= \sigma^l_{{AB}} \sigma_{{l}}^{AB}\qquad\quad \sigma_{{n}}^2:= \sigma^n_{{AB}} \sigma_{{n}}^{AB}\qquad\quad \sigma_{{n}}\cdot \sigma_{{l}}:=\sigma^n_{{AB}}\sigma_{{l}}^{{AB}}\qquad\quad
    \tilde \kappa=\frac{\tilde{\mathcal{D}}_{v}\eta}{\eta}+\frac{\partial_{r}V}{2}\,.
\end{equation}

\paragraph{Vector equations.} We list, respectively $\mathcal{E}_{nA}=0, \mathcal{E}_{lA}=0$,
\begin{subequations}
\begin{align}
&
    \frac{1}{\sqrt{{\cal G}}}\partial_{r}(\sqrt{{\cal G}}\beta_{A})+\frac{(D-3)}{(D-2)}\partial_{A}(\eta\theta_{n})-\tilde{\nabla}^{B}(\eta {\sigma^n_{AB}})=0\label{eom-nA}\\
&  
    \tilde{\mathcal{D}}_{v}\beta_{A}+\frac{V}{2}\partial_{r}\beta_{A}+\tilde{\nabla}_{A}\Big(\frac{\partial_{r}V}{2}+\frac{(D-3)}{(D-2)}\theta_l\Big)+\frac{\eta}{2}\frac{\theta_{n}\tilde{\nabla}_{A}V}{(D-2)}
   \nonumber \\  &\hskip 2cm 
    -\frac{\theta_l\tilde{\nabla}_{A}\eta}{\eta}-\tilde{\nabla}^{B}\sigma^l_{AB}+\frac{\eta}{2}\sigma^n_{AB}\tilde{\nabla}^{B}V=0\label{eom-lA}
\end{align}
\end{subequations}
where
\begin{equation}
    \beta_A = \frac{1}{2\eta} \left( g_{{AB}} \partial_r U^B +\partial_A \eta \right)= -\tilde {\mathcal{H}}_{A}  + \frac{\partial_{A}\eta}{\eta},
\end{equation}
\paragraph{Symmetric-traceless tensor equation.} The final set of equations is
\begin{multline}\label{eom-AB-trace-less}
    \frac{1}{\eta}\partial_{r}\sigma^l_{AB}+2\sigma^n_{(A}{}^{C}\sigma^l_{B)C}+\beta_A\beta_B+\tilde{\nabla}_{(A}\beta_{B)}-\frac{1}{2}\theta_l\sigma^{\text{\tiny $n$}}_{AB}-\frac{1}{2}\frac{(D-6)}{(D-2)}\theta_n\sigma^{{l}}_{AB}-\frac{1}{2}\tilde{R}_{AB}\\
    +\frac{1}{(D-2)}\Big(\frac{1}{2}\tilde{R}-\beta_A\beta^A-\tilde{\nabla}_A\beta^A\Big)g_{AB}=0\,.
\end{multline}

We now analyse the above equations assuming Taylor expandability in $r$ for the transverse metric.  Separating the transverse metric into its determinant $\mathcal G$ and a unimodular metric $\tilde \gamma_{AB}$, \eqref{eom-nn} implies that coefficients of expansion of the determinant corresponding to orders strictly bigger than one are specified in terms of the unimodular metric $\tilde \gamma_{AB}$ and lower orders of the determinant.  The two unspecified coefficients are encoded in $\Omega,\Theta_n$ in the conventions of section \ref{sec:solution space}.  Equations \eqref{eom-nA} and \eqref{eom-ln}  fix the radial dependence of $U^A, V$ respectively up to $2((D-2)+1)$ co-dimension one  functions. These can be  encoded in $\mathcal U^A,\Upsilon^A,\kappa$ and the leading order of $V_0$ can be put to zero, enforcing that ${\cal N}$ is a null surface. We hence have fixed all the radial dependence of the metric.

Using the $r$-component of contracted Bianchi identity, $\nabla_\mu(\mathcal E_{r\nu}g^{\mu\nu})=-\frac12 \partial_r\left( g^{AB}\right)\mathcal E_{AB}$, one deduces that only the leading order of \eqref{eom-AB-trace} has to be imposed. This corresponds to \eqref{Thetan-EoM}. Moreover, equations \eqref{eom-ll},\eqref{eom-lA} reduce to the Raychaudhuri \eqref{EoM-Raychaudhuri} and Damour \eqref{EoM-Damour}  equations. The remaining equations, \eqref{eom-AB-trace-less}, constrain the evolution of $\tilde \gamma_{AB}$, except for its zeroth order component. 

To summarize, the results derived in section \ref{sec:solution space} also apply to the case of Gaussian null-like gauge \eqref{G-F-M-01} with a Taylor expandable transverse surface.

\section{On covariant phase space} \label{appen:CPSF}

\begin{figure}[htb]
\def \L {3.0}
    \centering
\begin{tikzpicture}
  \draw[thick,red] (-0.8*\L,-0.8*\L) coordinate (b) -- (0.9*\L,0.9*\L) coordinate (t);
  \draw[black,thick,->] (0.83*\L,0.83*\L)--(0.84*\L,0.84*\L); \draw[black,thick,->] (0.85*\L,0.85*\L)--(0.86*\L,0.86*\L);
  \filldraw [black] (0*\L,0.0*\L) circle (2pt); \draw[blue] (-0.02*\L,0.01*\L) node[left] (scrip) {\small{${\cal N}_{v_1}$}}; 
  \filldraw [black] (0.45*\L,0.45*\L) circle (2pt); \draw[blue] (0.43*\L,0.46*\L) node[left] (scrip) {\small{${\cal N}_{v_2}$}};
  \draw[thick,, dashed, green] (-0.45*\L,-0.47*\L)  -- (0.45*\L,0.43*\L); \draw[black] (0.25*\L,0.23*\L) node[below, rotate=45] (scrip) {\textcolor{green}{\small{$\Sigma_2$}}};
  \draw[black] (0.83*\L,0.88*\L) node[left, rotate=45] (scrip) {{$v$}};
  \draw[red] (-0.6*\L,-0.5*\L) node[left, rotate=45] (scrip) {\small{$r=0$}}; 
  \draw[red] (-0.6*\L,-0.7*\L) node[left, rotate=45] (scrip) {\small{${\cal N}$}};
  \filldraw [blue] (-0.45*\L,-0.45*\L) circle (1pt); \draw[blue] (-0.42*\L,-0.43*\L) node[left] (scrip) {\small{${\cal N}_{v_0}$}};
  \draw[brown] (1.2*\L,0.7*\L) node[left] (scrip) {\small{$r>0$}};
  \draw[brown] (0.45*\L,0.7*\L) node[left] (scrip) {\small{$r<0$}};
\draw[thick,, dashed, green] (0,0.02*\L)  -- (-0.45*\L,-0.43*\L); \draw[black] (-0.25*\L,-0.23*\L) node[above, rotate=45] (scrip) {\textcolor{green}{\small{$\Sigma_1$}}};
\end{tikzpicture}
\caption{Null boundary ${\cal N}$ and segments $\Sigma_1,\Sigma_2$ on it.}
\label{Fig:Sigma1-2-B}
\end{figure}
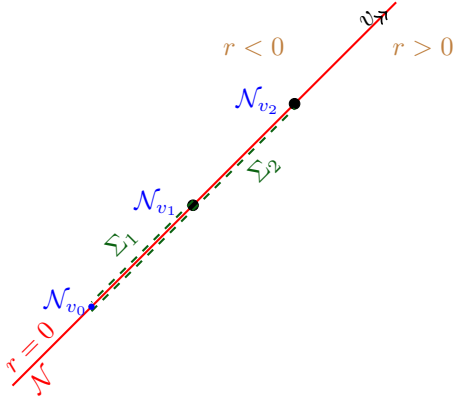

In this appendix, we briefly review how to associate a charge to a symmetry, focussing on cases where we have a null surface ${\cal N}$ as depicted in Fig.~\ref{Fig:Sigma1-2-B}. Then, we specialize the symplectic potential to Einstein gravity for the coordinate system adopted in \eqref{G-F-M-01}. 

\paragraph{Surface charge for a generic null surface.}
Starting from an action,
\be
S= \int \d{}^Dx\ {\cal L}
\ee
 the Lee-Wald symplectic current   $\omega_{\text{\tiny LW}}^{\mu}[\delta_1 g, \delta_2 g ; g]$ is defined as
\begin{equation}\label{LW-omega}
        \omega_{\text{\tiny LW}}^{\mu} [\delta g_1, \delta_ 2g ; g] = \delta_1 \Theta_{\text{\tiny LW}}^{\mu}[\delta_2g ; g] -\delta_2 \Theta_{\text{\tiny LW}}^{\mu}[\delta g_1 ; g]\qquad\qquad \delta {\cal L}\approx \partial_\mu \Theta_{\text{\tiny LW}}^{\mu}[\delta g ; g]
\end{equation}
where $\approx$ denotes on-shell equality. From the above one observes that the symplectic current is conserved on-shell,
\begin{equation}\label{conservation-01}
    \partial_\mu \omega_{\text{\tiny LW}}^{\mu}[\delta_1 g, \delta_2 g ; g] \approx 0\, .
\end{equation}
By virtue of the Poincar\'e lemma,  \eqref{conservation-01} implies
\begin{equation}\label{wdQ}
\omega_{\text{\tiny LW}}^\mu [\delta g, \delta_\xi g ; g]\approx \partial_\nu \mathcal{Q}_\xi^{\mu \nu}[\delta g; g]
\end{equation}
where $\mathcal{Q}_\xi^{\mu \nu}$ is a skew-symmetric tensor.

Consider the $r\geq 0$ part of spacetime bounded by a null boundary ${\cal N}$ and let $\Sigma_v$ be a section on ${\cal N}$ bounded between $v_0$ and $v_1$ or $v_2$, as depicted in Fig.~\ref{Fig:Sigma1-2-B}.

Let $\xi$ be the generator of a symmetry that generates variations $\delta_\xi g$ over the solution space, e.g., the ones discussed in section \ref{sec:NBS-generators}. The charge variation associated with the symmetry generator $\xi$ is defined as
\begin{equation}
\slashed{\delta} Q_{\xi} |_{\Sigma_v}:= \int_{\Sigma} \omega_{\text{\tiny LW}}^{\mu}[\delta g, \delta_\xi g ; g] \d{}^{D-1} x_{\mu}\,.
\end{equation}
Using \eqref{wdQ} and Stokes' theorem, one has 
\begin{equation}
\slashed{\delta} Q_{\xi} |_{\Sigma_v}\approx \int_{\partial\Sigma_v} \mathcal{Q}_\xi^{\mu \nu}[\delta g; g] \d{}^{D-2} x_{\mu}
\end{equation}
where $\partial \Sigma_v$ is the boundary of $\Sigma_v$. One then has 
\begin{equation}\label{conservation-02}
    \slashed{\delta} Q_{\xi} |_{\Sigma_2}-\slashed{\delta} Q_{\xi} |_{\Sigma_1} \approx \int_{\mathcal N_{v_1}}^{\mathcal N_{v_2}} \mathcal{Q}_\xi^{\mu \nu}[\delta g; g] \d{}^{D-2} x_{\mu}\, .
\end{equation}
In the limit $|v_2-v_1|\to 0$ this expression simplifies,
\begin{equation}
\frac{\d {}}{\d v} \left(\slashed{\delta} Q_{\xi}  -  \int_{\mathcal{N}_v} \mathcal{Q}_\xi^{vr} \d{}^{D-2} x\right)\approx 0\, .
\end{equation}
One can therefore consistently define the charge variation as a surface (co-dimension two) integral,
\begin{equation}\label{charge-variation-append}
 \slashed{\delta} Q_{\xi}  :=  \int_{\mathcal{N}_v} \mathcal{Q}_\xi^{vr} \d{}^{D-2} x
\end{equation}
at arbitrary values of $v$. Our derivation has bypassed any information about the bulk, about the asymptotia of spacetime, or the requirement of $\Sigma$ being a Cauchy surface.

The covariant phase space formalism reviewed above for the null boundary has inherent ambiguities of the symplectic potential that arise from using the Poincar\'e lemma on the spacetime ($W$) or on the phase space ($Y$), $\Theta^\mu\to\Theta^\mu+\partial_\nu Y^{\mu\nu}+\delta W^\mu$ \cite{Iyer:1994ys}. The $Y$-ambiguity affects the charge variation whereas the $W$-ambiguity is relevant for the boundary Lagrangian, the variational principle and could be relevant for the separation of the charge into integrable and flux parts \cite{Freidel:2021cbc}. We do not address these issues in our current work.

\paragraph{Explicit expression for the symplectic potential.} 
In our case we take ${\cal L}$ to be the Einstein--Hilbert Lagrangian, ${\cal L}=\frac{1}{16\pi G} \sqrt{-g} (R - 2 \Lambda)$, and get
\begin{equation}\label{symplpot}
    \Theta^{\mu}_{\text{\tiny LW}}[\delta g; g]=\frac{\sqrt{-g}}{16\pi \, G} \left( \nabla_\nu  (\delta g)^{\mu\nu} -\nabla^\mu (\delta g)_\nu^\nu \right)\,.
\end{equation}
The $r$-component of the symplectic potential, relevant to the charge analysis at any constant $r$ surfaces for the metric \eqref{G-F-M-01} or \eqref{metric-appendix-B}, is given by
\begin{multline}\label{Theta-r-generic-r}
 16 \pi G\, \Theta_{\text{\tiny LW}}^r= \frac{2}{D-2} \delta \Big[ n^r \sqrt{-g} \Big( \theta_l - \frac{\eta V}{2} \theta_{n}\Big) \Big] + \partial_\nu \Big[ 2 \sqrt{-g}  \Big(\delta n^{[r} l^{\nu]} - n^{[r} \delta l^{\nu]}\Big)\Big]\\
  \qquad +n^{r} \sqrt{-g} \Big[\Big( \sigma_{l}^{AB}-\frac{\eta V}{2}\sigma_{n}^{AB}\Big)\delta g_{{AB}} + 2\mathcal{H}_{\mu} \delta l^\mu +  \theta_{n} \, \delta (\eta V) \\
 + 2 \delta \Big( \kappa + \frac{D-3}{D-2} \Big(\theta_l - \frac{\eta V}{2}\,  \theta_n\Big) \Big) \Big]\,.
\end{multline} 

In particular, on the null surface $\mathcal N$ at $r=0$,  \eqref{Theta-r-generic-r} takes the form
\begin{multline}\label{LWsymplpot}
        16\pi G\Theta^{r}_{\text{\tiny LW}}[\delta{g};g]\big|_{{\mathcal{N}}} = -\Omega N^{AB}\delta{\Omega_{AB}} -2 \Omega \mathcal{H}_a \delta l^a + 2 \Big(\kappa +\frac{D-3}{D-2}\Theta_l\Big) \delta \Omega \\ 
        - 2 \delta \big(\Omega (\kappa +\Theta_l)\big) +\partial_a\Big(\frac{\Omega}{\eta}\delta ( \eta \,l^a)\Big)
\end{multline}
with $a$ labelling the boundary coordinates $v,x^A$. The quantities $(N^{AB},\mathcal{H}_a,\kappa+\frac{D-3}{D-2}\Theta_l)$ and $(\Omega_{AB},l^a,\Omega)$ are, respectively, the null equivalent of the usual stress energy tensor and the boundary metric that we have for timelike boundaries, and $\eta$ corresponds to a corner quantity related to the volume of the normal metric and its expansion \cite{Hopfmuller:2016scf,Hopfmuller:2018fni}. The first line in \eqref{LWsymplpot} contains the genuine flux, sourced by $N_{AB}$, and the non-conservation due to boundary sources $\Omega \mathcal{H}_{a}$, $\kappa+\frac{D-3}{D-2}\Theta_l$. In our analysis, we have left the dynamics of these sources unspecified.  
For completeness, we display the symplectic potential in terms of the charges $\mathcal P, \mathcal J^A, \Omega$ obtained in the Heisenberg slicing, 
\begin{multline}
        16\pi G\Theta^{r}_{\text{\tiny LW}}[\delta{g};g]\big|_{{\mathcal{N}}}= \partial_v (\Omega \delta  \mathcal{P})-\mathcal{J}_{A}\delta{\mathcal{U}^{A}}-\Omega N^{AB}\delta{\Omega_{AB}} +2  \Theta_l^{-1} N_{AB} N^{AB} \, \delta \Omega \\
        -2 \delta \Big( \Omega \, \Theta_l^{-1} \, N_{AB} N^{AB}  +\frac{\Theta_l \Omega}{D-2} \Big)+\partial_a(\Omega\delta l^a+ \Omega  l^a \delta \mathcal{P}) \,.
\end{multline}

We close this appendix with the remark that the second line in \eqref{LWsymplpot} involves terms that may be respectively absorbed into $Y$- and $W$-ambiguities of the symplectic potential. This point will be further explored elsewhere.

\section{Other families of genuine slicing} \label{appen:another-geuine-slicing}
In section \ref{sec:genuine-slicing} we worked through a one-parameter family of genuine slicings. This example already shows that genuine slicings are not unique. Here, we showcase two other families of such slicings and the associated algebras. 

\paragraph{Intermediate family.} Starting from the thermodynamic slicing in section \ref{sec:thermoslicing}, consider the following change of slicing
\begin{equation}\label{hat-slicing-10}
    \hat{W}=W-\Gamma T\qquad\qquad
    \qquad {\hat{Y}}^A={Y}^A +T \mathcal{U}^A\qquad\qquad \hat{T}^{(s)}=e^{-s \mathcal{P}}\Omega\Theta_l  T
\end{equation}
where $s$ is a real number and $\mathcal{P}$ is defined in \eqref{cal-P--def}. Using the adjusted bracket, one can deduce the algebra of null boundary symmetries for the intermediate slicing, 
\begin{equation}\label{NBS-hat-algebra}
    [\xi(  {\hat{T}}^{(s)}_1, {\hat{W}}_1, {\hat{Y}}_1^A), \xi( {\hat{T}}^{(s)}_2,  {\hat{W}}_2, {\hat{Y}}_2^A)]_{{\text{adj. bracket}}}=\xi(  {\hat{T}}^{(s)}_{12}, {\hat{W}}_{12}, {\hat{Y}}_{12}^A)
\end{equation}
where 
\begin{subequations}
    \begin{align}
        & {\hat{T}}_{12}^{(s)}=s({\hat{W}}_1 {\hat{T}}_2^{(s)}-{\hat{W}}_2 \hat{T}_1^{(s)})+\bar{\nabla}_A(\hat{T}_2^{(s)}\hat{Y}_1^A-\hat{T}_1^{(s)}\hat{Y}_2^A)\label{T12s}\\
        & {\hat{W}}_{12}={\hat{Y}}_1^A\bar{\nabla}_A{\hat{W}}_2-{\hat{Y}}_2^A\bar{\nabla}_A{\hat{W}}_1\\
        & {\hat{Y}}^A_{12}={\hat{Y}}_{1}^{B}\bar{\nabla}_B{\hat{Y}}_2^A-{\hat{Y}}_2^B\bar{\nabla}_B{\hat{Y}}_1^A
    \end{align}
\end{subequations}
for all $s \in \mathbb{R}$. At a given $v$, the above algebra is $\mathcal{A}^{(s)}_2 \inplus$ Diff$({\cal N}_v)$, where  Diff(${\cal N}_v)$ is generated by $Y^A$, $\mathcal{A}^{(s)}_2$ is generated by $\hat{T}^{(s)}$, and $\hat{W}$ is an algebra of the form 
\be
 [{\hat{T}}^{(s)}, {\hat{T}}^{(s')}]\sim 0\qquad \qquad [{\hat{T}}^{(s)}, {\hat{W}}]\sim s {\hat{T}}^{(s)}\qquad\qquad [\hat{W}, {\hat{W}}]\sim 0\,.
\ee 
As we see, $\hat{W}$ is a scalar under Diff$({\cal N}_v)$ whereas $\hat{T}^{(s)}$ is in a scalar density representation of Diff$({\cal N}_v)$, and \eqref{T12s} implies that $\mathcal{A}^{(0)}_2$ is the abelian $u(1)\oplus u(1)$ algebra.

The charge variation in the intermediate slicing \eqref{hat-slicing-10} reads as
\begin{equation}\label{charge-variation-s-slicing}
     \slashed{\delta} Q_{\xi}= \frac{1}{16\pi G} \int_{{\cal N}_v} \d{}^{D-2} x \left({\hat{W}}\delta\Omega+{\hat{Y}}^{A}\delta\Upsilon_{A}+{\hat{T}}^{(s)}\delta{\mathcal{P}}_{(s)}-{\hat{T}}^{(s)} e^{s\mathcal{P}}\ \Theta_l^{-1} N^{AB}\delta\Omega_{AB} \right)
\end{equation}
with
\begin{equation}\label{expansion-charge-aspect}
 {\mathcal{P}}_{(s)}=
  \begin{cases} 
    \frac{1}{s}\,e^{s\mathcal{P}}=\frac1{s} \left(\frac{\eta}{\Theta^2_{l}}\right)^s & \text{if } s \neq 0 \\
   \mathcal{P}       & \text{if } s =0 \, .
  \end{cases}
\end{equation}
One can split the charge variation \eqref{charge-variation-s-slicing} into integrable and flux parts using the MB method, yielding
\begin{align}
     \hat{Q}^{\text{I}}_{\xi} &= \frac{1}{16\pi G} \int_{{\cal N}_v} \d{}^{D-2} x \left({\hat{W}} \Omega+\hat{Y}^A \Upsilon_A+\hat{T}^{(s)}\, \mathcal{P}_{(s)} \right) \\
     \hat{F}_{\xi}(\delta g) &= 
     \frac{1}{16\pi G} \int_{{\cal N}_v} \d{}^{D-2} x \,\ \Omega T\, N_{AB}\,\delta\Omega^{AB}\, .
\end{align}
As we see explicitly, in the intermediate slicing the charges are integrable in the absence of genuine flux. This means it is indeed an example for a genuine slicing.

Also, the intermediate slicing keeps the Weyl charge aspect $\Omega$ and angular momentum aspect $\Upsilon_A$ the same as in the thermodynamic slicing. We dub the charge associated with rescaled $v$-translations, ${\cal P}_{(s)}$ ``expansion aspect'', since ${\cal P}_{(s)}$ for $s<0$ is proportional to $\Theta_l^{-2s}$, \emph{cf.}~\eqref{expansion-charge-aspect}. This charge vanishes for $\Theta_l=0$, see section \ref{sec:non-expanding}.

The transformations laws for the intermediate slicing
\begin{subequations}
    \begin{align}
        &\delta_{\xi}\Omega={\hat{T}}^{(s)} e^{s\mathcal{P}}+\bar{\nabla}_A(\Omega{\hat{Y}}^{A})\\
        &\delta_{\xi}\Upsilon_{A}=-{\hat{T}}^{(s)}\bar{\nabla}_A {\mathcal{P}}_{(s)} +\Omega\bar{\nabla}_A{\hat{W}}+\mathcal{L}_{{\hat{Y}}}\Upsilon_{A}-2\bar{\nabla}^{B}\left({\hat{T}}^{(s)} e^{s\mathcal{P}} \Theta_l^{-1}N_{AB}\right)\\
        &\delta_{\xi} {\mathcal{P}}_{(s)}=-( \delta_{s,0}+ s {\mathcal{P}}_{(s)}) \, \hat{W} +{\hat{Y}}^{A}\bar{\nabla}_A {\mathcal{P}}_{(s)}+\frac{2{\hat{T}}^{(s)} e^{{2}s\mathcal{P}}}{\Omega\Theta_l^{2}}N_{AB}N^{AB}
    \end{align}
\end{subequations}
together with the definition of the MB yield 
\begin{equation}\label{BT-Bracket-hat-s}
     \left\{{\hat{Q}}^{\text{I}}_{\xi_{1}},{\hat{Q}}^{\text{I}}_{\xi_{2}}\right\}_{\text{\tiny{MB}}}={\hat{Q}}^{\text{I}}_{[\xi_{1},\xi_{2}]_{{\text{adj. bracket}}}}+ \delta_{s,0} \, {\hat{K}}^{(0)}_{\xi_{1},\xi_{2}}
\end{equation}
where 
\begin{equation}
    {\hat{K}}^{(0)}_{\xi_{1},\xi_{2}}= \frac{1}{16\pi G} \int_{{\cal N}_v} \d{}^{D-2} x \left({\hat{W}}_{1}{\hat{T}}^{(0)}_{2}-{\hat{W}}_{2}{\hat{T}}^{(0)}_{1}\right).
\end{equation}
Hence, the charge algebra reads as
\begin{subequations}
    \begin{align}
        &\{\Omega(v,x),\Omega(v,x')\}=0\\
        &\{{\mathcal{P}}_{(s)}(v,x),{\mathcal{P}}_{(s')}(v,x')\}=0\\
        &\{\Omega(v,x),{\mathcal{P}}_{(s)}(v,x')\}=16\pi G\left(s {\mathcal{P}}_{(s)}(v,x)+\delta_{s,0}\right)\delta^{D-2}\left(x-x'\right)\\
        &\{\Upsilon_A(v,x),\Upsilon_B(v,x')\}=16\pi G\left(\Upsilon_{A}(v,x')\partial_{B}-\Upsilon_{B}(v,x)\partial'_{A}\right)\delta^{D-2}\left(x-x'\right)\\
        &\{\Upsilon_A(v,x),\Omega(v,x')\}=-16\pi G \,\Omega(v,x) \partial'_{A}\delta^{D-2}\left(x-x'\right)\\
        &\{\Upsilon_A(v,x),{\mathcal{P}}_{(s)}(v,x')\}=16\pi G\left(-{\mathcal{P}}_{(s)}(v,x)\partial'_{A}-{\mathcal{P}}_{(s)}(v,x')\partial_{A}\right)\delta^{D-2}\left(x-x'\right)\,. \label{Ups-P}
    \end{align}
\end{subequations}
The above algebra, as expected and by construction, is of the form $\mathcal{C}^{(s)}_2 \inplus$ Diff$({\cal N}_v)$. The $\mathcal{C}^{(s)}_2$ part is generated by $\Omega,{\mathcal{P}}_{(s)}$ and Diff$({\cal N}_v)$ by $\Upsilon_A$. This algebra is not of a direct sum form. Nonetheless, it may be brought to a direct sum form upon another change of slicing, as discussed in the main text in section \ref{sec:genuine-slicing}. This explains why we refer to this slicing as intermediate.

\paragraph{Another family.}
Just as yet-another example of a genuine slicing, consider alternatively the following change of slicing
\begin{equation}
   \hat{W}=W-\Gamma T+\frac{2}{D-2} \Theta_l T\qquad\qquad \hat{T}^{(s)}=\frac{\Omega\Theta_l}{\Xi^s} T
\end{equation}
where 
\begin{equation}
    \Xi := 
    \eta\Theta_l^{-2}\Omega^{-\frac{2}{D-2}}\,.
\end{equation}
In this other slicing it is assumed that the hatted quantities are field-independent,  $\delta{\hat{W}}=\delta{\hat{T}}=\delta{\hat{Y}^A}=0$. The algebra of these symmetry generators is then
\begin{equation}\label{NBS-hat-algebra-2}
    [\xi(  \hat{T}^{(s)}_1, \hat{W}_1, \hat{Y}_1^A), \xi( \hat{T}^{(s)}_2,  \hat{W}_2, \hat{Y}_2^A)]_{{\text{adj. bracket}}}=\xi(  \hat{T}^{(s)}_{12}, \hat{W}_{12}, \hat{Y}_{12}^A)
\end{equation}
where
\begin{subequations}
    \begin{align}
        & \hat{T}_{12}^{(s)}=\hat{Y}_1^A\bar{\nabla}_A\hat{T}_{2}^{(s)}
        +\frac{D+2s-2}{D-2}\hat{T}_2^{(s)}\bar{\nabla}_A\hat{Y}^A_1
        +s\hat{W}_1\hat{T}^{(s)}_2 - (1\leftrightarrow 2)\\
        & \hat{W}_{12}=\hat{Y}_1^A\bar{\nabla}_A\hat{W}_2-\hat{Y}_2^A\bar{\nabla}_A\hat{W}_{1}\\
        & \hat{Y}^{A}_{12}=\hat{Y}_1^B\bar{\nabla}_B\hat{Y}_2^A-\hat{Y}_2^B\bar{\nabla}_B\hat{Y}_1^A\,.
    \end{align}
\end{subequations}
The above algebra is $\mathcal{A}_2 \inplus$ Diff$({\cal N}_v)$, where  Diff(${\cal N}_v)$ is generated by $Y^A$ and $\mathcal{A}_2$ by $\hat{T}^{(s)}$ and $\hat{W}$. As we see, $\hat{W}$ is a scalar under Diff$({\cal N}_v)$ whereas $\hat{T}^{(s)}$ is in a scalar  density  representation of Diff$({\cal N}_v)$. For the special case of $s=-\frac{D-2}{2}$, $\hat{T}^{(s)}$ is also a scalar under Diff$({\cal N}_v)$.

To explore what is the $\mathcal{A}_2$ part, we turn off the Diff$({\cal N}_v)$ part, where we remain with an algebra of the form $[T, T]\sim 0,\ [W,W]\sim 0,\ [T,W]\sim s T$. This algebra for $s\neq 0$ is closely related to a Heisenberg algebra (upon redefining $e^{T/s}$ as the new generator we get a Heisenberg algebra). For $s=0$ the algebra $\mathcal{A}_2$ is $u(1)\oplus u(1)$.

In this slicing the charge variation obtains
\begin{equation}
     \slashed{\delta} Q_{\xi}= \frac{1}{16\pi G} \int_{{\cal N}_v} \d{}^{D-2} x \left(\hat{W}\delta\Omega+\hat{Y}^{A}\delta\Upsilon_{A}+\hat{T}^{(s)}\delta\mathcal{P}_{(s)}-\hat{T}^{(s)}\Xi^s\Theta_l^{-1} N^{AB}\delta\Omega_{AB} \right)
\end{equation}
with
\begin{equation}\label{charges-3d}
 \mathcal{P}_{(s)}=
  \begin{cases} 
    \frac{1}{s}\,\Xi^s & \text{if } s \neq 0 \\
   \ln{\Xi}       & \text{if } s =0\,.
  \end{cases}
\end{equation}
Integrable and flux parts can be separated using the MB method, 
\begin{align}
     \hat{Q}^{(s)\text{I}}_{\xi}&= \frac{1}{16\pi G} \int_{{\cal N}_v} \d{}^{D-2} x \left(\hat{W} \Omega+\hat{Y}^{A} \Upsilon_{A}+\hat{T}^{(s)}\ \mathcal{P}_{(s)} \right)\\
     \hat{F}_{\xi}^{(s)}(\delta g)&= -\frac{1}{16\pi G} \int_{{\cal N}_v} \d{}^{D-2} x \,\hat{T}^{(s)}\Xi^s\Theta_l^{-1} N^{AB}\delta\Omega_{AB} \, .
\end{align}

The charge algebra may also be computed, yielding
\begin{equation}\label{BT-Bracket-hat-s-2}
     \left\{\hat{Q}^{\text{I}}(\xi_1),\hat{Q}^{\text{I}}(\xi_2)\right\}_{\text{\tiny{MB}}}=\hat{Q}^{\text{I}}([\xi_1,\xi_2]_{{\text{adj. bracket}}})+\hat{K}_{\xi_1,\xi_2}
\end{equation}
where
\begin{equation}
    \hat{K}_{\xi_{1},\xi_{2}}= \frac{1}{16\pi G} \delta_{s,0}\int_{{\cal N}_v} \d{}^{D-2} x\left[\hat{W}_{1}\hat{T}_{2}-\hat{W}_{2}\hat{T}_{1}-\frac{2}{D-2}\left(\hat{Y}_{1}^{A}\partial_{A}\hat{T}_{2}-\hat{Y}_{2}^{A}\partial_{A}\hat{T}_{1}\right)\right]\,.
\end{equation}
More explicitly, 
\begin{subequations}
    \begin{align}
        &\{\Omega(v,x),\Omega(v,x')\}=0\\
        &\{\mathcal{P}_{(s)}(v,x),\mathcal{P}_{(s)}(v,x')\}=0\\
        &\{\Omega(v,x),\mathcal{P}_{(s)}(v,x')\}=16\pi G\left(s\mathcal{P}_{(s)}(v,x)+\delta_{s,0}\right)\delta^{D-2}\left(x-x'\right)\\
        &\{\Upsilon_A(v,x),\Upsilon_B(v,x')\}=16\pi G\left(\Upsilon_{A}(v,x')\partial_{B}-\Upsilon_{B}(v,x)\partial'_{A}\right)\delta^{D-2}\left(x-x'\right)\\
        &\{\Upsilon_A(v,x),\Omega(v,x')\}=-16\pi G \Omega(x,v) \partial'_{A}\delta^{D-2}\left(x-x'\right)\\
        &\{\Upsilon_A(v,x),\mathcal{P}_{(s)}(v,x')\}=16\pi G\Big[-\mathcal{P}_{(s)}(v,x)\partial'_{A}-\frac{D+2s-2}{D-2}\mathcal{P}_{(s)}(v,x')\partial_{A}\nonumber\\ &\hspace{5.25cm}+\frac{2\delta_{s,0}}{D-2}\partial'_{A}\Big]\delta^{D-2}\left(x-x'\right)\,. \label{Ups-P-1}
    \end{align}
\end{subequations}
As we see, and as expected, $s=0, -\frac{D-2}{2}$ are special values. For $s=0$ the algebra is a semi-direct sum of Heisenberg and Diff$({\cal N}_v)$. The Heisenberg part $\Omega$ is a scalar and $\cal P$ a scalar density of weight $-1$ under Diff$({\cal N}_v)$. For $s=-\frac{D-2}{2}$ the quantities $\Omega$ and ${\cal P}$ fall into the same representation of Diff$({\cal N}_v)$.

\section{Kerr metric in Gaussian null coordinates} \label{appen:Kerr-Metric-GNC}
The Kerr black hole in Boyer--Lindquist coordinates is 
\begin{equation}\label{Kerr-metric}
    \d s^2=-\frac{\Delta}{\rho^2}(\d t-a\, \sin ^{2}\theta \d{} \hat{\phi})^2+\frac{\rho^2}{\Delta}\d{}\hat{r}^2+\rho^2 \d{}\theta^2+\frac{(\hat{r}^2+a^2)^2 \sin ^2\theta}{\rho^2}(\d{}\hat{\phi}-\frac{a}{\hat{r}^2+a^2}\d t)^2
\end{equation}
with
\begin{equation}
    \Delta=\hat{r}^2-2M\hat{r}+a^2\qquad\qquad \rho^2=\hat{r}^2+a^2 \text{cos}^{2}\theta
\end{equation}
where $M$ and $J=aM$ are mass and angular momentum of the black hole, respectively. The outer and inner horizon radii $r_{\pm}$ are given by the bigger and smaller roots $\Delta=0$,
\begin{equation}
    r_{\pm}=M\pm \sqrt{M^2-a^2}\,.
\end{equation}
This stationary axisymmetric black hole geometry has a Killing horizon at $\hat{r}=r_+$ generated by the Killing vector
\begin{equation}
    \xi_{\text{\tiny{H}}}=\partial_{t}+\Omega_{\text{\tiny{H}}}\partial_{\hat{\phi}}
\end{equation}
with horizon angular velocity $\Omega_{\text{\tiny{H}}}$ and surface gravity $\kappa_{\text{\tiny{H}}}$
\begin{equation}\label{KerrSurfGrav}
    \Omega_{\text{\tiny{H}}}=\frac{a}{a^2+r_{+}^2}\qquad\qquad \kappa_{\text{\tiny{H}}}=\frac{r_+-r_-}{2r_+(r_++r_-)}\,.
\end{equation}

In the Gaussian null coordinates with horizon located at $r=0$ \cite{Booth:2012xm},
\begin{equation}
    g_{\mu \nu}= g_{\mu \nu}^{(0)} + r\, g_{\mu \nu}^{(1)}+ \mathcal{O}(r^2)
\end{equation}
the Kerr metric reads
\begin{align}
    g_{vr}^{(0)} & =1 &  g_{\theta\theta}^{(0)} &=\rho_{+}^2 \\ 
    g_{\phi\phi}^{(0)}&=\frac{(r_{+}^{2}+a^2)^2\sin ^2\theta}{\rho_{+}^2} & 
  g_{vv}^{(1)}&=-2 \kappa_{\text{\tiny{H}}} \\  
  g_{v \theta}^{(1)}&= \frac{2 a^{2} \sin \theta \cos \theta}{\rho^2_+} &  g_{v \phi}^{(1)} &= \frac{a \sin ^{2} \theta}{ \rho^{{4}}_+ } \left[ \rho^2_+ (r_+ -r_-)+2 r_{+}  (r_{+}^{2}+a^{2}) \right]\\
       g_{\theta \theta}^{(1)} &= \frac{2 r_{+} (r_{+}^{2}+a^{2})}{ \rho^2_+} & g_{\phi \phi}^{(1)}&= \frac{2 r_{+} (r_{+}^{2}+a^{2})^{2}  \sin ^{2} \theta}{ \rho^6_+} \left[ 2 \rho^2_+ -(r_{+}^{2}+a^{2})\right]\\
        g_{\theta \phi}^{(1)}&= - \frac{2 a^{3} (r_{+}^{2}+a^{2}) \sin^{3} \theta \cos \theta}{\rho^4_+}\,. &&
\end{align}
In term of our solution space variables of section \ref{sec:solution space}, we have
\begin{align}
    \eta&=1 & \Omega_{\theta\theta}&=\rho_{+}^2 & \Omega_{\phi\phi}&=\frac{(r_{+}^{2}+a^2)^2\sin^2\theta}{\rho_{+}^2} & \Omega&=(r_{+}^{2}+a^2)\sin\theta\\
      \kappa&=\kappa_{\text{\tiny{H}}} & \Gamma&=-2\kappa_{\text{\tiny{H}}} & \Theta_l&={\cal U}^\theta={\cal U}^\phi=0 & \Theta_n&=-\frac{2 r_+}{\rho_+^2}
\end{align}      
and
\begin{align}
           \Upsilon^{\theta}&=-\frac{2r_+^3\Omega^2_{\text{\tiny{H}}} (r_++r_-)^3}{\rho^{{4}}_+} \sin^2 \theta \cos \theta &
          \Upsilon^{\phi}&=-2a\Big(\kappa_{\text{\tiny{H}}}+\frac{r_+}{\rho_{+}^2}\Big)\sin\theta\\
        \lambda_{\theta\theta}&=-\frac{ r_{+}^2 (r_{+}+r_-)}{ \rho^2_+} & 
    \lambda_{\theta\phi}&= \frac{a^{3} (r_{+}^{2}+a^{2}) \sin^{3} \theta \cos \theta}{\rho^4_+} \\
        \lambda_{\phi\phi}&=-\frac{r_{+} (r_{+}^{2}+a^{2})^{2}  \sin ^{2} \theta}{ \rho^6_+} \left[ 2 \rho^2_+ -(r_{+}^{2}+a^{2})\right]\,. &&
\end{align}
The surface charges \eqref{chargenonexp} for the Kerr black hole are
\begin{align}
    S_{\text{\tiny{Kerr}}} &=4\pi\, Q(0,\tilde{W},0)=\frac{\pi}{G} M r_+\, \tilde{W}_{0}(v)\, \qquad \qquad
    J_{\text{\tiny{Kerr}}}=- Q(0,0,\tilde{Y}^{A})=\frac{aM}{G}\, \tilde{Y}^{\phi}_{0}(v)
\end{align}
where $\tilde{W}_{0}(v)=\tilde{W}(v,x=0)$ and $\tilde{Y}^{\phi}_{0}(v)=\tilde{Y}^{\phi}(v,x=0)$.

\addcontentsline{toc}{section}{References}
\bibliographystyle{fullsort.bst}
\bibliography{reference}

\providecommand{\href}[2]{#2}\begingroup\raggedright
\endgroup

\end{document}